# Frequent Query Matching in Dynamic Data Warehousing

Charles H. Goonetilleke, J. Wenny Rahayu, and Md. Saiful Islam

*La Trobe University, Melbourne, Australia*
∗ *Corresponding author. Tel.: +61-3-9479-1282; fax: +61-3-9479-3060; e-mail: w.rahayu@latrobe.edu.au*

## ABSTRACT

With the need for flexible and on-demand decision support, Dynamic Data Warehouses (DDW) provide benefits over traditional data warehouses due to their dynamic characteristics in structuring and access mechanism. A DDW is a data framework that accommodates data source changes easily to allow seamless querying to users. Materialized Views (MV) are proven to be an effective methodology to enhance the process of retrieving data from a DDW as results are pre-computed and stored in it. However, due to the static nature of materialized views, the level of dynamicity that can be provided at the MV access layer is restricted. As a result, the collection of materialized views is not compatible with ever-changing reporting requirements. It is important that the MV collection is consistent with current and upcoming queries. The solution to the above problem must consider the following aspects: (a) MV must be matched against an OLAP query in order to recognize whether the MV can answer the query, (b) enable scalability in the MV collection, an intuitive mechanism to prune it and retrieve closely matching MVs must be incorporated, (c) MV collection must be able to evolve in correspondence to the regularly changing user query patterns. Therefore, the primary objective of this paper is to explore these aspects and provide a well-rounded solution for the MV access layer to remove the mismatch between the MV collection and reporting requirements. Our contribution to solve the problem includes a Query Matching Technique, a Domain Matching Technique and Maintenance of the MV collection. We developed an experimental platform using real data-sets to evaluate the effectiveness in terms of performance and precision of the proposed techniques.

*Keywords: Frequent Matching, Dynamic Data Warehousing,*

## 1. Introduction

Businesses collect large amounts of data from transactions in order to gain insight into their performance and growth. Decisions must be made based on this data to support the ever-changing needs of customers and business requirements. In present days' the data warehouse has become a popular tool for decision making among businesses. Traditional data warehouses (DW) are subject-oriented, time variant, non-volatile and integrated data repositories that are capable of providing decision support as data is kept in an aggregated form [16]. Commonly, the data from transactional data sources are extracted and transformed into an aggregated form via the Extract-Transform-Load (ETL) process to be stored in the data warehouse and then onwards users can either execute On-Line Analytical Processing [3, 5, 18, 19] (OLAP) queries on it or create smaller data marts from it. Several data sources may have to be integrated to create the data warehouse. The query results can then be used for analytical purposes, for example to identify customer trends, so that business owners or management can make important business decisions based on this information.

In order to retrieve information from a data warehouse, users can either directly submit an OLAP query or use an existing Materialized View (MV), which contains the pre-computed result for a particular query. This type of information is mainly used for compiling reports for decision making purposes. Reports may be produced on a daily, weekly, monthly and/or yearly basis, and results from one or more queries would be required. Therefore, storing a number of MVs for this type of frequently used queries would be beneficial as query results need not be computed at runtime. This would mean that the utilization of MVs would reduce query execution time, as opposed to direct querying that requires expensive database operations such as joins, grouping, nesting to be performed at runtime. For example, if the cost of accessing the query result is $C_1$, then directly querying would have a total cost of $C_1 + O_2$, where $O_2$ is the cost of performing the aforementioned operations in DW. Contrary to this, the total cost of using an MV to get the result is just $C_1$, as the result is already computed. Hence MVs are an effective solution to faster reporting.

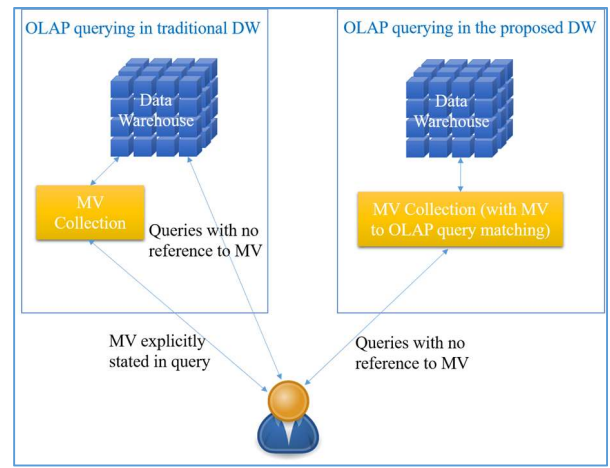

**Figure 1. OLAP querying in traditional DW and proposed DW**

The presence of an MV collection in a data warehouse is advantageous to the reporting process. However, the following issues exist from the user perspective:

- Users are required to determine which MV(s) can answer their query; and
- Users must explicitly reference the MV(s) in their queries.

This means that users are required to know intricate details about the collection and re-write their queries, if they wish to utilize MVs as illustrated in left hand side of Figure 1. Furthermore, from a DW perspective, the following issues are observed:

- The MV collection must be intuitive so that majority of the queries are answered using MVs;
- When the MV collection is large, finding a match by a linear scan is time consuming; and

- MVs must be up-to-date in order to provide non-stale results.

In general, the MVs in the collection would be required to undergo changes as a consequence of the changes occurring in reporting requirements. The objective is to provide results for majority of queries using MVs, and therefore identification of most and least frequent queries is important. The frequency of usage of each MV may change at any given time, therefore MVs may require to be added or removed from the collection. For example, a particular report that used $MV_1$ may change so that it requires $MV_2$ instead. $MV_1$ would no longer be in use so would have to be removed. Then $MV_2$ must be added to the collection. Since traditional DWs are not able to automate these operations, the database administrator (DBA) would have to analyze the frequency of usage of each MV and determine which MV(s) should be remained/removed/added. In most DWs, the collection of MVs is expected to be quite large, hence finding a match for a query can be a time-consuming task. To ensure the results stored in MVs are up-to-date, these MVs must be updated whenever a change occurs in the underlying DW. For a large collection of MVs, this update process can take a lot of time as results must be re-computed. Due to the static characteristics of the MV collection, traditional DW impose lesser usability of MVs to its users.

Existing *dynamic data warehouses* (DDW) aim to provide dynamicity to statically designed and implemented traditional data warehouses (DW). For example, the approaches presented by the authors of the articles [17, 22, 26] focus more on the design aspect of a DDW. In addition to that, concepts such as Slowly Changing Dimensions [9, 16, 19] (SCDs), Multi-Version Data warehouse [10] (RTDW) are other techniques that can transform the static DW to one that can accommodate structural and data changes seamlessly. The concept of Slowly Changing Measures [14, 15] (SCMs) is another recent approach to DDW design, which aims to enhance the ability of a DW to adapt to measurement function changes. Though dynamicity in the design of a DDW is an important aspect, the access to a DDW through dynamically maintained MV collection might play a major role in improving its usability. The works in [13, 21] propose techniques to keep the MVs up-to-date in relation to the changes in the data sources. However, there exists no work on dynamic maintenance of MV collection based on user queries and thereafter, answering the user queries by implicitly matching them to the appropriate MVs in the collection.

In this paper, we aim to introduce dynamicity in the MV collection of DDW as illustrated on the right-hand side of Figure 1. However, the introduction of dynamicity in the MV collection is a non-trivial problem and there are several issues that need to be addressed. Firstly, a technique to match an MV and an OLAP query must be defined. Next, in order to support the scalability of the MV collection, a faster access mechanism needs to be incorporated within it. Finally, the MV collection must be maintained by removing inactive MVs and adding new high demand MVs through identification of new query patterns, thus majority of user queries would be answered using the pre-computed results in MVs. To address the above issues, our contribution consists of the followings:

(1) **Query Matching**: we propose a novel technique to match an OLAP query to an MV in the collection;

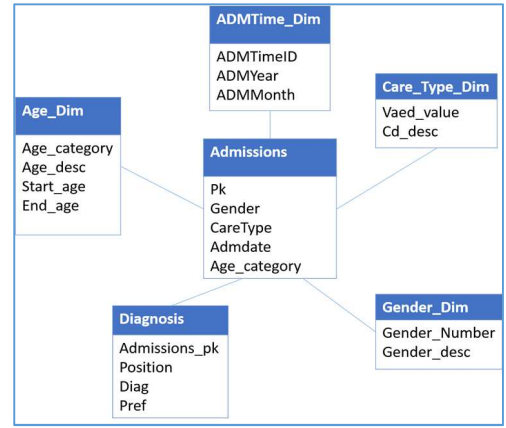

**Figure 2. An example of DW Star Schema**

(2) **Domain Matching**: we also propose a technique to add scalability to the MV collection for faster retrieval of matching MVs;

(3) **Maintenance of MV Collection**: we propose a technique to maintain the timeliness of the MV collection in relation to user query patterns; and

(4) We have also implemented each of the above techniques and performed detailed experimentation to demonstrate the effectiveness and efficiency of our approach.

With the above key points brought to context, the remainder of this paper is organized into the following sections: Section 2 provides the preliminaries. Section 3 formally presents the problem studied in this paper and captures the key challenges that need to be addressed. Section 4 presents our framework of solution. Section 5 explores the notion of matching an MV to an OLAP query and presents our proposed Query Matching technique. This section also presents our Domain Matching technique to address the scalability issue of matching an OLAP query to an MV in the collection. Section 6 presents our approach to maintain the timeliness of the MV collection based on user query patterns. Section 7 evaluates the effectiveness and performance of all of our proposed approaches by experimenting with real dataset. Finally, Section 8 concludes the findings of this paper.

## 2. Preliminaries

### 2.1. Data Warehouse

Data warehouses are central data repositories that contain consolidated data from one or more data sources for analytical and decision making purposes. The data stored in a DW takes a multi-dimensional form, as information can be retrieved from different points of view; for example, time, location, category. Conceptually, a DW is designed using a *star schema* [11]. It consists of *fact* tables and *dimension* tables. A fact is a measurement or metric of a transaction or collection of transactions. The granularity of the DW is described as the depth of information provided by the fact table(s). High granularity is similar to transactional level of data while low granularity data is summaries usually via grouping. The fact column(s) of the fact table contains this data and other columns are foreign keys to data that describe the facts. The foreign keys point to tables known as dimension tables, which contains descriptive information about each row. An illustration of a data warehouse star schema with fact (Admissions) and dimension tables (Age_Dim, ADMTime_Dim, Care_Type_Dim, Gender_Dim, Diagnosis) is given in Figure 2.

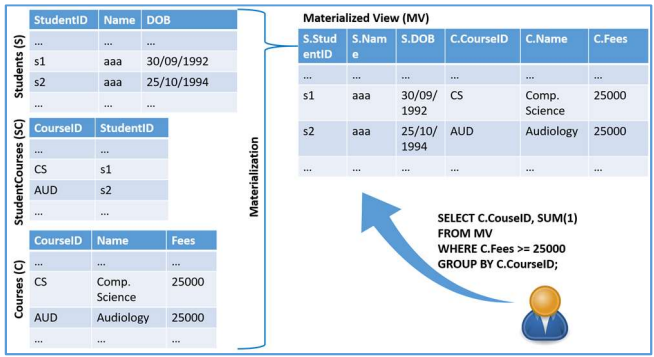

**Figure 3. An example of a Materialized View**

*2.2. Materialized View*

A materialized view (MV) contains the cached result of a particular OLAP query. In contrast to Views in databases as well as DWs, which contain only the definition of a query, an MV consists of the query and also, the query result. MVs are used primarily to store results for large queries for reusability and performance benefits. After materializing a query result in an MV, users can query them by including the reference(s) in the FROM clause of queries. Figure 3 illustrates the creation and querying of an MV.

As retrieving the result of OLAP queries in large scale DWs can be time consuming, MVs are usually used for faster response in practice.

### 3. Problem Definition

For a DDW with an MV collection, two types of OLAP queries can be presented. The first type are queries that contain explicit references to MV(s) – results are directly obtained from the collection. The second type are regular queries that have references to tables of the DW and some of these queries can be answered using MV(s). In this paper, we target queries of the second type that can be processed by accessing the MV(s) although MV references are not explicitly given in the user query. Therefore, we state our problem formally as follows:

*Definition 3.1*: Given an OLAP query Q (S, F, W, G), where S is the set of projection columns, F is the set of tables, W is the set of selection conditions and G is the set of group by attributes, we want to find an appropriate match for the tables of F in the collection of materialized views M such that $Q^F = Q^M$, where $Q^F$ denotes the result of evaluating Q on the list of tables F in DDW and $Q^M$ denotes the result of evaluating Q in the collection M. Otherwise, decide whether $Q^F$ should be stored in M to answer the same or queries that can be answered from $Q^F$.

The solution to the above problem has two major challenges to be addressed. Firstly, we need to determine whether M can answer Q. To do so, we need to compare the materialized views in M to an OLAP query Q. Since the two entities are different, the derivation of a comparison algorithm is not a trivial problem. The main challenge regarding this is to provide a common platform for both entities so that the comparison would be accurate. In this paper, we define what determines the match between $Q^F$ and $Q^M$, called *query matching*, which is discussed in Section 5. However, the number of MVs in M can be large, and if so, the query matching process might take longer time than evaluating Q directly in DDW. That is, if $T_M$ is the total time taken to find a matching MV in M and then retrieve the result from M, and $T_{DDW}$ is the time taken for direct querying in DDW, then the condition $T_M < T_{DDW}$ must be satisfied. To satisfy this requirement, we propose an efficient *domain matching* technique to prune the search space so that only MVs that are most likely to have a match with the user query are used in *query matching*, which is discussed in detail in Section 6.

Secondly, with regards to determining whether or not to store $Q^F$ as a materialized view, the main issue is the limited space available in the MV collection. Having an intuitive and highly usable list of MVs is important as it will ensure most user queries will be answered using MVs in the collection. To add an MV into the collection, we need to consider how often users will require it. To delete an existing MV and make space for $Q^F$, we need to determine if the deleted MV is no longer used by the users. If the latter case is true we also need to determine if users may require the MV again on a later occasion. Furthermore, the contents of each MV must be updated to ensure integrity in the results provided by MVs. These issues of *MV maintenance* in M are addressed in detail in Section 7 of this paper.

### 4. The Proposed Framework

The proposed framework of solution is illustrated in Figure 4, which consists of two important modules: (a) *Query Matching* (QM) – matches an incoming OLAP query Q to the corresponding query $Q_{MV}$ of an MV in the MV collection; this module also includes a submodule called *Domain Matching* (DM), which reduces the number of $Q_{MV}$s that are needed to be checked against Q before performing the actual match; and (b) MV Maintenance (MM) – maintains the MV collection so that high frequency queries can have matching MVs for faster response.

*4.1. Query Matching*

The QM module in our framework takes the incoming user query and the stored MV collection M as inputs and compares each MV in M to the user query Q. The matching algorithm then attempts to determine if the result expected by Q is derivable by the MVs in M. Since an MV is essentially a stored query result, the algorithm first extracts its corresponding query $Q_{MV}$, where $Q_{MV}$ is a query of a MV in the collection M. Both $Q_{MV}$ and Q are then transformed into a common format to do the actual matching and determine whether Q can be answered from the corresponding MV. The detail of the matching algorithm is given in Section 5.

QM improves the usability of the MV collection as users are not required to explicitly reference MVs in their queries. However, comparing each $Q_{MV}$ to Q can cause QM to be slower with large MV collections. Hence, we propose *domain matching* (DM) as a pruning technique to reduce the search space for QM. In DM we identify and remove MVs with a potential lower match to Q at an earlier stage of the matching process thereby the next stage (QM) receives a subset of the collection with higher probability for a match. A key advantage of this approach is that if an entirely new query is entered and none of the MVs match, then the pruning stage recognizes this as early as possible. The resulting subset for the next stage then becomes *null* and so the query to MV matching operations need not be performed. As a consequence, computation time and resources for an entirely new query can be saved. This is discussed in detail in Section 5.

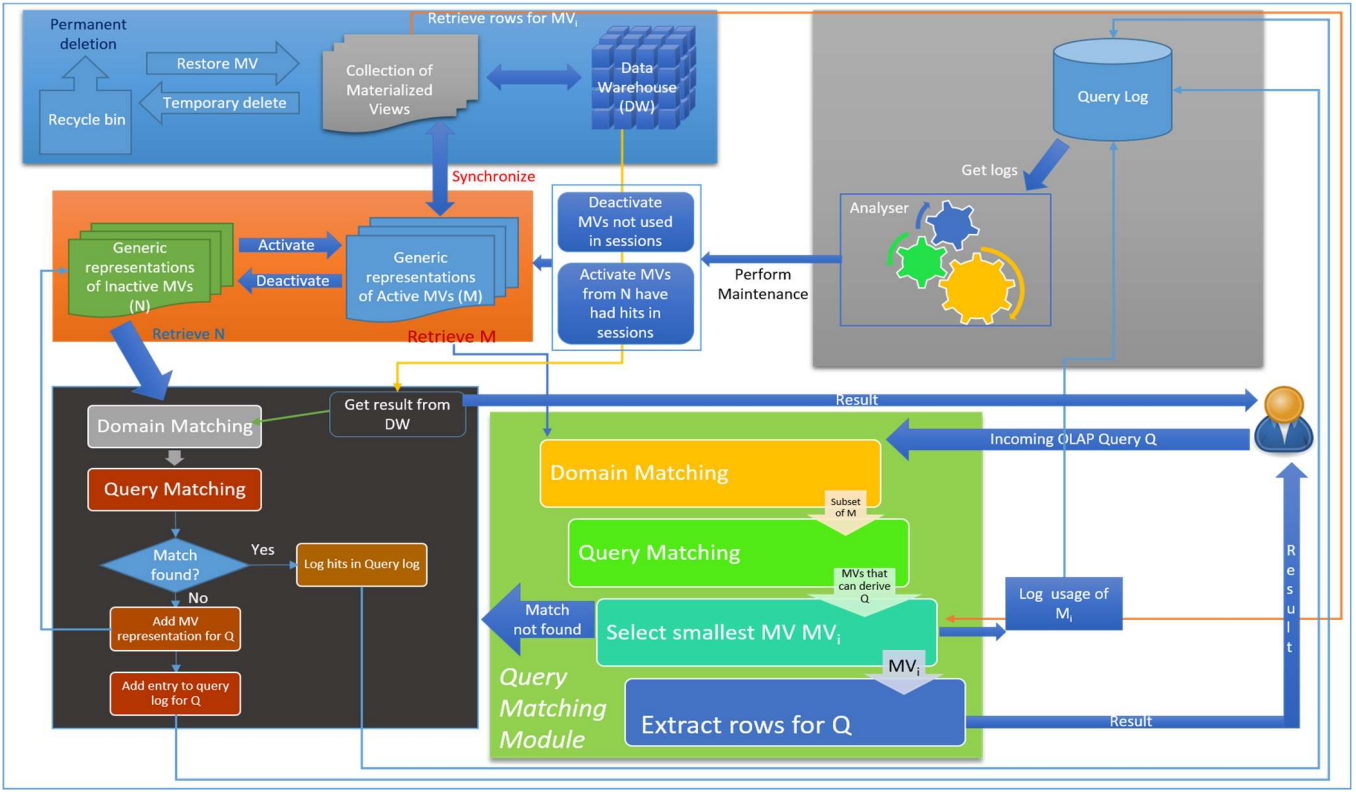

**Figure 4. Framework of Solution**

*4.2. MV Maintenance*

The frequency of the usage of the MVs in the collection M can change over time. In order to increase the success rate of DM and QM the collection of materialized views in M must be updated regularly to match user query patterns. In addition to that, the results stored in MVs must be up-to-date with the DW. To satisfy these requirements, we perform the following actions on M and more details of each type of action are given in section 7: (a) maintenance of a query log – keep track of the hits of each $Q_{MV}$ for each OLAP session, (b) create new MVs for OLAP queries that are most frequent – a mechanism to determine which queries are frequent, (c) temporary-delete MVs– move MV out of M to secondary collection N, (d) permanent-deletion – delete outdated MVs from N, and (e) keep track of DW changes for view maintenance – a mechanism to monitor the DW and update the MVs when the underlying DW changes. Our framework runs the above as an isolated process so that the users can run their OLAP queries seamlessly.

**5. Query Matching**

Matching the OLAP query Q to the queries of the MVs in M, i.e., $Q_{MV}$s, is not trivial. Existing approaches for finding the similarity between these two types of queries are not applicable. In this section, first we review existing approaches to identify their strengths and weaknesses for the above. We then propose a novel two step technique for matching Q to $Q_{MV}$: (a) firstly, we represent both Q and $Q_{MV}$ by a unified structure, and (b) then, we apply a set of heuristics rules to do the actual matching. As the matching process requires to compare each $Q_{MV}$ in M to Q, we also propose *domain matching* as a pruning technique to reduce the search space, i.e., $Q_{MV}$s in M, for the input query Q.

*5.1 Existing Approaches*

Determining the similarity between two entities has been vastly explored in the literature, especially in the area of information retrieval [6]. In a DW environment, determining similarity between OLAP queries have been explored in areas such as query recommendation [1, 8] and query personalization [2]. The following are some common models used to represent queries and thereafter, determine the similarity in the literature:

- String of the interpreted SQL statement [24] : the string of a SQL query with another is a straightforward methodology of determining the similarity as edit-distance measure such as Levenshtein Distance [7] can be used. Here we can either compare the entire string or split the query into sections and compare each one independently and finally take the summation of the distances. This method is not suitable for matching OLAP queries as queries such as the following would be determined as non-similar, although the result of them are identical.

    $Q_1$ = "select id, description from fact_table";
    $Q_2$ = "select description, id from fact_table";

- Vectors of features [4] : contain vectors of features to store a score for each characteristic in the queries and cosine similarity is then used to determine the similarity. The authors of the article [12] propose an algorithm known as SIMCHECK to perform this operation. In the first stage of the algorithm, the feature vectors are compared to determine if the number of tables, sum of the table degrees, and the sum of the join index and predicate counts are equal. In the next phase structural features are compared to determine similarity. The main advantage of this approach is that if the queries are found to be dis-similar in the first phase, continuation into the second phase is aborted and so computation resources and time are saved. Though SIMCHECK can be used to prune the number of $Q_{MVS}$ to be compared with Q in our problem settings, it requires complicated feature vector construction and thereafter Hamming distance calculation with each $Q_{MV}$ in M, which is not efficient.
- Represent query as a graph [23] : The database schema is used to create graphs for two queries and then a graph comparison algorithm is used to find similarity. However, this approach can introduce unnecessary complexity in our problem settings. Representing a query as a tree would be a more suitable approach, which has been explored in our approach.

### 5.2 Our Approach

In this section, we present our approach for matching an OLAP query Q to the queries $Q_{MV}$ in M. To achieve this, we firstly define a common/unified structure for representing both, Q and $Q_{MV}$, known as *OLAP query tree*. The matching algorithm then takes these representations as its inputs to determine the match between them. We also outline the limitation of our approach on queries involving aggregate functions except summation. Finally, we present a novel technique to prune non-promising MVs in M, which can reduce the number of candidate $Q_{MVS}$ to be matched for Q significantly.

#### 5.2.1 Unified Structure for $Q_{MV}$ and Q

Considering the limitations of existing approaches for matching queries in our problem settings, we identified that the following criteria must be considered for matching Q to $Q_{MV}$:

(a) Evaluate $Q_{MV}$ and Q from bottom to top – as query execution engine of a database generates execution plan using the bottom-up evaluation;
(b) Compare each type of fragment – compare SELECT, FROM, WHERE, GROUP BY conditions of $Q_{MV}$ and Q;

To satisfy these requirements we propose a variation of *relational algebra query tree* as depicted in Figure 5, known as *OLAP Query Tree* (OQT). As each segment of a query such as the SELECT, FROM, WHERE and GROUP BY are placed in separate nodes, each fragment can be evaluated and then compared in isolation. With reference to [25], the structure of a OQT can be formally defined as follows:

An OQT $Q = (V_d, V_o, V_r, E_q, f_e)$, where:

(1) $V_d$ and $V_o$ are finite sets of nodes, known as, Data nodes and Operator nodes, respectively. And we define $V_q$ as the complete set of nodes, where $V_q = V_d \cup V_o$.
(2) If $u_1 \in V_d$ is a table T and can be represented as $T \subseteq \prod_{a \in A} D_a$, such that for each $f \in T$ has a domain A and each $f(a) \in D_a$, then $V_r \subseteq A$ known as Result columns.
(3) $E_q \subseteq \{(u_1, u_2) \mid u_1, u_2 \in V_o\} \cup \{(u_1, u_2) \mid u_1 \in V_o, u_2 \in V_d\}$ is a finite set of edges and $(V_q, E_q)$ is a directed graph.
(4) $f_e$ is a function for nodes $u_1 \in V_o$ and $u_2 \in V_d$ that define the number of children it can have, i.e. $f_e(u_1) \in [1,2]$ and $f_e(u_2) = 0$

All nodes in OQT can be categorized into two types as, Operator Node and Table node. Operator nodes contain either a distinct, projection, selection, group by, having, join or Cartesian product operation, where each node produces a number of columns as the result. The queries Q and $Q_{MV}$ are represented in the OQT in their un-optimized form i.e., each type of condition is placed in one node only. For example, all selection conditions are placed in a single node. All nodes except for Table nodes are allowed to have 1-2 children, where only Join and Cartesian product nodes have exactly 2 child nodes. The types of child nodes each node can have are summarized in Table 1.

| Node | Number of children | Allowed Nodes as Children |
|---|---|---|
| Projection | 1 | All nodes |
| Group by | 1 | Projection, Join, Cartesian Product, Table |
| Selection | 1 | Projection, Join, Cartesian Product, Table |
| Cartesian Product | 2 | Projection, Join, Cartesian Product, Table |
| Table | 0 | None |

**Table 1. Possible Child Nodes of OQT Nodes**

Using OQT to represent the structure of a query is beneficial as fragments of queries can be compared with each other. The OQT is set to contain nodes for projection, selection, group by and cartesian product operator only. The OQT can also be easily extended to include the structure of inner queries. For the purpose of query matching in dynamic data warehousing, we add the following restrictions to the OQT:

(a) Distinct and Having nodes constrict the result of an MV to be very specific and hence the usage of such MVs would be low.
(b) Inner and outer joins were excluded from the OQT as these can be broken down as cartesian products followed by one or more selection conditions.
(c) All selection conditions are placed in one selected node.
(d) All projection conditions are placed in one projection node.
(e) All OLAP queries must contain one or more group by conditions as OLAP queries are used to retrieve aggregated results.

| Row | Attr1 | Attr2 | Attr5 |
|---|---|---|---|
| 1 | | | |
| 2 | | | |
| ... | | | |
| n | | | |

Pre-computed result of MV1: Contains all rows and only the columns attr1, attr2 and attr5

| Row | Attr1 | Attr3 |
|---|---|---|
| 1 | | X |
| 2 | | X |
| ... | | |
| m | | X |

Result of Q1: Columns attr1 and attr3 and rows where attr3 value is X

| Row | Attr1 | Attr2 |
|---|---|---|
| 1 | | |
| 2 | | |
| ... | | |
| p | | |

Result of Q2: Columns attr1 and attr2 and rows where attr3 value is X

**Figure 5. MV Result Comparison with Query Result**

(f) All selection conditions are placed in one selected node.
(g) All projection conditions are placed in one projection node.
(h) All OLAP queries must contain one or more group by conditions as OLAP queries are used to retrieve aggregated results.

### 5.2.2 Matching a $Q_{MV}$ to OLAP query Q

Let us consider the following example OLAP queries $Q_3$ and $Q_4$, and the $Q_{MV1}$ in M as given as follows:

$Q_3$ = SELECT attr1, attr3 FROM table1 WHERE attr3='X';

$Q_4$ = SELECT attr1, attr2 FROM table1 WHERE attr3='X';

$Q_{MV1}$ = SELECT att1, attr2, attr5 FROM table1;

The two queries $Q_3$ and $Q_4$ are similar with respect to their tables and the where conditions, but only differ in terms of one attribute in the select statement. Therefore, it can be stated that $Q_3$ and $Q_4$ are closely related. A comparison between $Q_4$ and $MV_1$ would yield a different kind of result. Although $Q_4$ is not closely related to the query of $MV_1$, it is certainly a valid match to the view's pre-computed results. $MV_1$ contains results for all rows of table1 along with only the three columns, attr1, attr2 and attr5. $Q_4$ requires a result where it contains the columns attr1 and attr2 and only rows where attr3 is equal to 'X'. This means that the result of $Q_4$ is a subset of the result of $MV_1$ and therefore $Q_{MV1}$ is a match to $Q_2$. Moreover, matching $Q_3$ with $Q_{MV1}$ returns a mismatch. $MV_1$ is able to provide all the rows required by $Q_3$ but since $MV_1$ does not contain the column attr3, it cannot provide a complete answer to $Q_3$. If we use similarity measures for queries only and not for data, this information would not be captured. Figure 6 illustrates the result of $MV_1$ and the query results of $Q_3$ and $Q_4$.

Therefore, to find a match between a $Q_{MV}$ and Q, the matching algorithm must identify if the result of Q is already a subset of the result of $Q_{MV}$. The result produced by evaluating each node after the final cartesian product of $OQT_Q$, which is the OQT representation of Q, must be a subset of the result produced after evaluating the counterpart node of the OQT representation of $Q_{MV}$, i.e., $OQT_{MV}$. The evaluations proceed in a bottom-up manner as depicted in Figure 7.

For the $OQT_Q$ and $OQT_{MV}$ illustrated in Figure 7 the output of columns produced by the final cartesian product are the same. Hence, we can proceed to match the remaining nodes. Assume that 100 rows are produced at $n_1$ and $n'_1$ and both have same tables, and 50 rows are produced at $n_2$. The number of rows at $n'_2$, with comparison to $n_2$, will be much less than 50 as it has an extra selection condition. Therefore, the rows produced at $n'_2$ is a subset of the rows produced at $n_2$. Hence, at points $n_2$ and $n'_2$, it can be stated that $OQT_Q$ is derivable using $OQT_{MV}$. At $n_3$ aggregation is performed under two attributes whereas at $n'_3$ aggregation is performed under one attribute and both have SUM(R.c) as the aggregate function. That is, the level of granularity of the result for $OQT_Q$ is lower than the corresponding result present in $OQT_{MV}$. At $n_3$ and $n'_3$, a similar scenario occurs as depicted in Figure 8 and Figure 9. As the grouping attribute R.b is not present in $OQT_Q$, further summation is required with respect to only R.a. The sum of R.c, are already available at point 3 and therefore, the result required at point II of $OQT_Q$ is derivable using the result in $OQT_{MV}$ at point 3 by further aggregating its result to the R.a attribute. Next it is observed that $n'_4$ contains a subset of the projection columns of $n_4$ and the aggregate functions specified at $n_4$ and $n'_4$ are the same; hence it is a match.

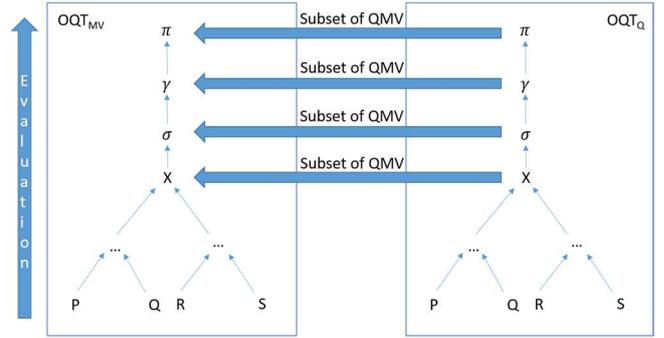

**Figure 6. Comparing OQT$_Q$ to OQT$_{MV}$**

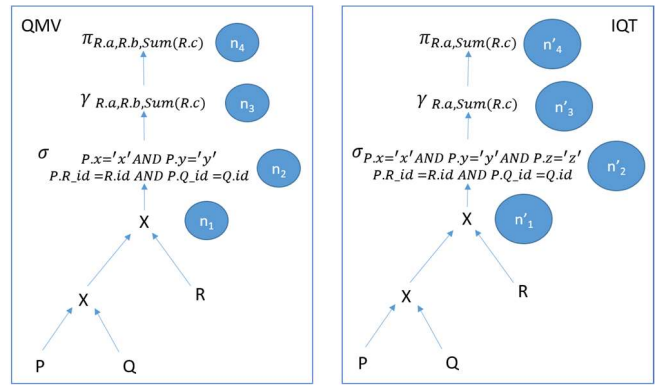

**Figure 7. Example for OQT$_Q$ to OQT$_{MV}$ matching**

| R.a | R.b | Sum(R.c) |
|---|---|---|
| PA23-250 | ZGF-516 | 20 |
| PA23-250 | AFR-987 | 30 |
| C-90A | AFF-124 | 10 |
| C-90A | MNB-467 | 80 |

**Figure 8 Group by two attributes for sum**

| R.a | Sum(R.c) | | R.a | Sum(R.c) |
|---|---|---|---|---|
| PA23-250 | 20 | | PA23-250 | 50 |
| PA23-250 | 30 | | | |
| C-90A | 10 | | y C-90A | 90 |
| C-90A | 80 | | | |

**Figure 9 Group by one attribute for sum**

Query matching begins at the leaf nodes which contain tables, the $OQT_{MV}$ and $OQT_Q$ must be equivalent in terms of their tables. Then the matchings are performed in the order

Selection ($\sigma$), Group by ($\gamma$) and Projection ($\pi$). The group by evaluation is split into two where firstly the Group by Attributes ($\gamma$) are compared, then Aggregate Functions ($\omega$) are compared. For all the following matchings between a OQT$_{MV}$ and OQT$_Q$, Levenshtein Distance [7] is used for string comparison.

*A. Matching Selection Conditions*

In order for an OQT$_Q$ and OQT$_{MV}$ to be similar with respect to their selection conditions, all conditions of OQT$_{MV}$ must be in OQT$_Q$. This means that the data in OQT$_{MV}$ can be reused by OQT$_Q$ and if it contains more conditions it would retrieve a subset of the data from OQT$_{MV}$. For each selection condition of OQT$_Q$ that match with that of OQT$_{MV}$, a score of 1 will be given and if no match the score will be 0. The sum of these scores divided by the number of conditions in OQT$_Q$ gives a value between [0,1]. This should be between equal to 1 as otherwise the OQT$_Q$ and the OQT$_{MV}$ are not a match with respect to their selection conditions.

Given an input query OQT$_Q$ which consists a set of conjunctive selection conditions $S_i = \{c1_i, c2_i, \ldots, cn_i\}$ and materialized view query OQT$_{MV}$ which consists of a set of conjunctive selection conditions $S_j = \{c1_j, c2_j, \ldots, cn_j\}$

Function $f(c_n)$ where $c_n \in S_j$, is defined such that if $c_n \in S_i$ then $f(c_n) = 1$ and if $c_n \notin S_i$ then $f(c_n) = 0$.

Sum of selection condition matches (S$_C$) = $\sum_{n=1}^{m} f(c_n)$, where m is $|S_j|$. Therefore, the match between the selection conditions OQT$_Q$ and OQT$_{MV}$ is $\sigma = \frac{S_C}{|S_j|}$, where $0 \leq \sigma \leq 1$.

The above applies only if both $S_1$ and $S_2$ have at least one element. If not the following applies,

if $S_1 = \phi$ and if $S_2 = \phi$, then $f(c_n) = 1 \Rightarrow \sigma = 1$

if $S_1 = \phi$ and if $S_2 \neq \phi$, then $f(c_n) = 0 \Rightarrow \sigma = 0$

if $S_1 \neq \phi$ and if $S_2 = \phi$, then $f(c_n) = 1 \Rightarrow \sigma = 1$

*B. Matching Group By Attributes*

In order for the OQT$_Q$ and the OQT$_{MV}$ to match with respect to their group by attributes, all group by attributes of OQT$_Q$ must be in OQT$_{MV}$ as well. For each attribute in OQT$_Q$ that match with that of OQT$_{MV}$, a score of 1 will be given and if no match the score will be 0. The sum of these scores divided by the number of group by attributes in OQT$_Q$ gives a value between [0,1]. This should be equal to 1 as otherwise the OQT$_Q$ and the OQT$_{MV}$ are not a match with respect to their group by attributes.

Given an OQT$_Q$ which consists a set of group by attributes $S_i = \{g1_i, g2_i, \ldots, gn_i\}$ and a OQT$_{MV}$ which consists of a set of group by attributes $S_j = \{g1_j, g2_j, \ldots, gn_j\}$

Function $f(g_n)$ where $g_n \in S_i$, is defined such that if $g_n \in S_j$ then $f(g_n) = 1$ and if $g_n \notin S_j$ then $f(g_n) = 0$

Sum of group by attribute matches (S$_G$) = $\sum_{n=1}^{m} f(g_n)$, where m is $|S_i|$

Therefore, the match between the grouping attributes of $S_i$ and $S_j$ is $\gamma = \frac{S_G}{|S_i|}$, where $0 \leq \gamma \leq 1$.

The above applies only if both $S_1$ and $S_2$ have at least one element. If not the following applies,

if $S_1 = \phi$ and if $S_2 = \phi$, then $f(c_n) = 1 \Rightarrow \gamma = 1$

if $S_1 = \phi$ and if $S_2 \neq \phi$, then $f(c_n) = 0 \Rightarrow \gamma = 0$

if $S_1 \neq \phi$ and if $S_2 = \phi$, then $f(c_n) = 1 \Rightarrow \gamma = 1$

*C. Matching Aggregate Functions*

In order for a OQT$_Q$ and a OQT$_{MV}$ to match with respect to their aggregate functions, all functions of OQT$_Q$ must be in OQT$_{MV}$ as well. For each function in OQT$_Q$ that match with that of OQT$_{MV}$, a score of 1 will be given and if no match the score will be 0. The sum of these scores divided by the number of group by functions in OQT$_Q$ gives a value between [0,1]. This should be equal to 1 as otherwise the query and the materialized view are not a match with respect to their aggregate functions.

Given an OQT$_Q$ which consists a set of group by attributes $S_i = \{ag1_i, ag2_i, \ldots, agn_i\}$ and a OQT$_{MV}$ which consists of a set of group by attributes $S_j = \{ag1_j, ag2_j, \ldots, agn_j\}$

Function $f(ag_n)$ where $ag_n \in S_i$, is defined such that if $ag_n \in S_j$ then $f(ag_n) = 1$ and if $g_n \notin S_j$ then $f(ag_n) = 0$

Sum of aggregate function matches (S$_A$) = $\sum_{n=1}^{m} f(ag_n)$, where m is $|S_i|$

Therefore, the match between the aggregate functions of $S_i$ and $S_j$ is $\omega = \frac{S_A}{|S_i|}$, where $0 \leq \omega \leq 1$.

The above applies only if both $S_1$ and $S_2$ have at least one element. If not the following applies,

if $S_1 = \phi$ and if $S_2 = \phi$, then $f(c_n) = 1 \Rightarrow \omega = 1$

if $S_1 = \phi$ and if $S_2 \neq \phi$, then $f(c_n) = 0 \Rightarrow \omega = 0$

if $S_1 \neq \phi$ and if $S_2 = \phi$, then $f(c_n) = 1 \Rightarrow \omega = 1$

The last case would happen only if there was a syntax error in the query. If OQT$_Q$ has grouping and the OQT$_{MV}$ does not, then the group by attribute match calculation would produce 0 as its result.

*D. Matching Projection Attributes*

In order for an OQT$_Q$ and a OQT$_{MV}$ to be similar with respect to their projection columns, all projection columns of OQT$_Q$ must be in OQT$_{MV}$ as well. For each column OQT$_Q$ that match with that of OQT$_{MV}$, a score of 1 will be given and if no match the score will be 0. The sum of these scores divided by the number of columns in OQT$_Q$ gives a value between [0,1]. This should be equal to 1 as otherwise the OQT$_Q$ and OQT$_{MV}$ are not a match with respect to their projection columns.

Given an OQT$_Q$ which consists a set of projection columns and $S_i = \{p1_i, p2_i, \ldots, pn_i\}$ and a OQT$_{MV}$ which consists of a set of projection columns $S_j = \{p1_j, p2_j, \ldots, pn_j\}$.

Function $f(c_n)$ where $c_n \in S_i$, is defined such that if $c_n \in S_j$ then $f(c_n) = 1$ and if $c_n \notin S_j$ then $f(c_n) = 0$

Sum of projection condition matches (S$_P$) = $\sum_{n=1}^{m} f(c_n)$, where m is $|S_i|$

Therefore, the match between the projections columns of the queries OQT$_Q$ and OQT$_{MV}$ is $\pi = \frac{S_P}{|S_i|}$, where $0 \leq \pi \leq 1$.

The above applies only if both $S_1$ and $S_2$ have at least one element. If not the following applies,

```
Algorithm 1 OQT_MV to OQT_Q matching algorithm
Input: σ, γ, ω, π. The values are in the range [0,1]
Output: The score
 1: score = 0
 2: if σ equals 1 then
 3:     score += 1
 4: end if
 5: if γ equals 1 then
 6:     score += 1
 7: end if
 8: if ω equals 1 then
 9:     score += 1
10: end if
11: if π equals 1 then
12:     score += 1
13: end if
14: return score
```

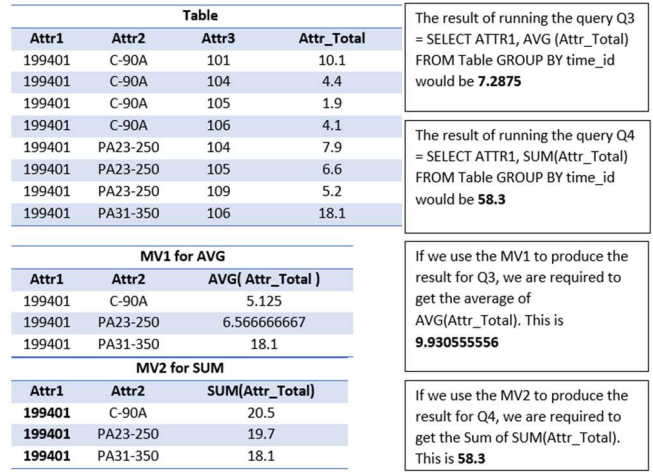

**Figure 10. Behavior of sum and avg in levels of aggregation**

if $S_1 = \phi$ and if $S_2 = \phi$, then $f(c_n) = 1 \Rightarrow \omega = 1$

if $S_1 = \phi$ and if $S_2 \neq \phi$, then $f(c_n) = 0 \Rightarrow \omega = 0$

if $S_1 \neq \phi$ and if $S_2 = \phi$, then $f(c_n) = 1 \Rightarrow \omega = 1$

Each pair of fragments being matched produces a score in the range [0,1] as the number of elements checked is divided by the cardinality of the set. If the score produce by a certain fragment matching is 1, it means that the $OQT_Q$ fragment is derivable using the corresponding fragment of the $OQT_{MV}$. Therefore, if the score is less than 4 then it indicates that one fragment matching did not succeed and hence the algorithm will not proceed any further. As a result, the $OQT_Q$ is not derivable using the $OQT_{MV}$. If the score is 4, it means that each fragment of the $OQT_Q$ is derivable using the corresponding $OQT_{MV}$ fragment, hence the result of $OQT_Q$ is derivable using $OQT_{MV}$ and thereby the $OQT_Q$ and $OQT_{MV}$ are a match. The above is pseudocoded in Algorithm 1.

### 5.2.4  Limitations in aggregate functions

For OLAP queries the aggregation functions are SUM, AVG, COUNT, MAX and MIN. It is important to note that being able to derive the result to an OLAP query using an MV is restricted to only the SUM aggregate function. For example, consider Figure 10. The derived average using the MV is not equal to the actual value expected by the query. Hence AVG cannot be used as an aggregate function for a $OQT_{MV}$ or an $OQT_Q$. Similarly, this same behavior is displayed with the aggregate functions COUNT, MAX and MIN. As granularity decreases, only the SUM aggregate function preserves its associativity and thereby the derivation of the result of a query using an MV must be restricted to the SUM aggregate function.

### 5.2.5  Optimization via Domain Matching

Comparison of each $Q_{MV}$ in M to Q can be a very time-consuming operation. To reduce the search space of M, we propose *domain matching* (DM) as an optimization technique. Assume the following scenarios take place where OQTs for Q and $Q_{MV}$ are being evaluated at the final cartesian products, denoted by × in Figure 11:

Scenario 1: Tree 1 is $OQT_{MV}$ and Tree 2 is $OQT_Q$. The $OQT_{MV}$ does not have enough information to derive the result of the $OQT_Q$ as it does not have the table Q in its result.

**Figure 11. Domain matching scenarios**

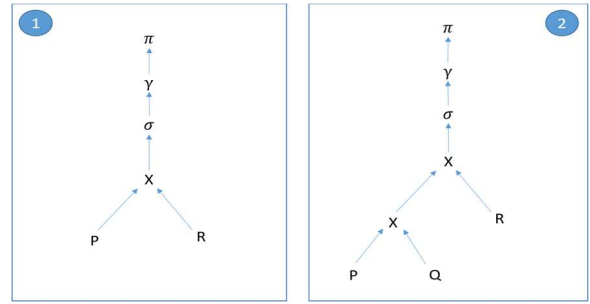

**Figure 12. Domain matching hash table example**

Scenario 2: Tree 1 is $OQT_Q$ and Tree 2 is $OQT_{MV}$. Since table Q is in the $OQT_{MV}$, one or more of its attributes may either be used as a grouping attribute or a selection attribute. Therefore, the result required by the $OQT_Q$ is not derivable by the $OQT_{MV}$.

As both Q and $Q_{MV}$ are OLAP queries, a group by clause will be present. Therefore, at least one attribute of each table specified in the query will either be used as a group by attribute or as a selection condition's attribute. Moreover, all group by attributes are present in the projection along with one or more aggregate functions. As a result of these constraints, the tables in $OQT_{MV}$ and $OQT_Q$ must be equivalent. We define the comparison between $OQT_{MV}$ and $OQT_Q$ in terms of their tables as *domain matching*. A hash table is used to further optimize the domain matching process. The type of keys used for hashing are the table names and the value for each table are the materialized views that references it. An example of a hash table used for domain matching is illustrated by Figure 11. The algorithm extracts the tables of $OQT_Q$ and extracts the values for those from the hash table as lists. The intersection of these lists is the $OQT_{MV}$ that have the same tables as $OQT_Q$ and therefore will be proceeding to the query matching stage. The above domain matching technique is pseudocoded in Algorithm 2.

---

**Algorithm 2** Domain Matching
**Input**: Q: Input query, hash_table: inverted list containing tables names as keys and $Q_{MVS}$ as values of keys
**Output**: Collection of $Q_{MVS}$
**Uses**: lookup (hash_table, key): Looks up a hash table and returns the value,
insert(map, key, value): Insert a key with value to a map. Replace value if key exists, insert(list, value): insert into list the given value, size(collection): number of item in collection
**Variables**: mvs: a map with table name as key and list of MVs as value, counts: a map with MV as key and an integer as value

```
1: T = Table names referenced in Q
2: result = empty set of MVs
3: for all t ∈ T do
4:     value := lookup(hash_table, t)
5:     if value equals ∅ then
6:         insert (mvs, t, value)
7:     else
8:         value := empty list of MVs
9:         insert (mvs, t, value)
10:    end if
11: end for
12: for all t ∈ T do
13:    temp := empty set of MVs
14:    temp <- lookup (mvs, t)
15:    for all q ∈ temp
16:        count := an integer
17:        count <- lookup (counts, q)
18:        if count equals null then
19:            insert (counts, q, 1)
20:        else
21:            count = count + 1
22:            insert (counts, q, count)
23:        end if
24:    end for
25: end for
26: for all (key, value) ∈ counts do
27: if size(value) equals size(T) then
28:    insert (result, value)
29: end if
30: return result
```

### 5.3 The Overall Query Matching Algorithm

The overall process of finding a match to Q from M begins by retrieving all active $Q_{MVS}$ to be passed to domain matching. The subset of $Q_{MVS}$ obtained after applying domain matching are then passed to query matching to determine a match for Q. If several $Q_{MVS}$ are a match to Q, the $Q_{MV}$ with the smallest number of rows is selected. Then the rows of the corresponding MV are retrieved from the DW and the relevant conditions of Q are applied to this result set before sending the query result to the user. In the event of no matching $Q_{MV}$ is found for Q, the system retrieves the query results directly from the DW.

---

**Algorithm 3** Overall Query Matching
**Input:** Q: incoming OLAP query
**Output:** rows from DB table

```
1: result = DB table rows
2: QMVs = retrieve all active QMVs from M
3: QMVs = perform domain matching on QMVs
4: if QMVs.size = 0 then
5:     result = retrieve result for Q from DW
6: else
7:     QMVs = perform query matching on QMVs
8:     if QMVs.size = 0 then
9:         result = retrieve result for Q from DW
10:    else
11:        QMV = select smallest QMV from QMVs
12:        result = retrieve rows for QMV
13:        result = apply relevant conditions of Q on result
14:    end if
15: end if
16: return result
```

### 5.4 Summary

In this section, we defined the structure and algorithm for matching the query of an MV to an input query Q. We introduced the *OLAP query tree* as the formal representation for $Q_{MV}$ and Q. The query matching algorithm takes a pair of these representations as its inputs and determines if the result expected by Q is within $Q_{MV}$. As an optimization technique *domain matching* is proposed. In the domain matching, all $Q_{MVS}$ with the same tables as Q are retrieved from an inverted list. The output of *domain matching* is used as input by query matching technique to find $Q_{MV}$(s) that can derive Q. The overall query matching algorithm utilized the above-mentioned techniques to determine a match and then retrieves the result from the DW, which is pseudocoded in Algorithm 3.

## 6 Maintenance

In order to have a set of MVs that can provide answers to a high proportion of user queries with up-to-date results, the collection MVs in M must be maintained. The maintenance process must consider all sources of change and then update M. According to the authors of [20] the development of a data warehouse (DW) must satisfy the information needs of the end-users and the available data sources. Based on this hypothesis, we propose a novel approach to maintain the MV collection by considering the query patterns of the user along with the changes of the sources in the underlying DW.

### 6.1 Analyse User Query Patterns

In order to dynamically add or remove an MV from the collection based on user query patterns, an *Analyzer* component is incorporated into our technique. The analyzer runs as an independent process, monitoring the usage of existing $Q_{MVS}$ in M and predict their upcoming usage. Upon completion of an analysis cycle, $Q_{MVS}$ are either added to/removed from M. In this paper, we define the changes done on M as *Slowly Changing Views* (SCV). As a consequence of SCVs, a materialized view will undergo a number of stages in its life cycle. Let us consider a DDW with the materialized views $MV_2$, $MV_3$ and $MV_4$. Assume that queries $Q_5$, $Q_6$ and $Q_7$ are the possible queries that users can enter; and assume that the result of $Q_5$ is derivable by $MV_2$ and the answer of $Q_6$ is derivable by MV$_3$. Q$_7$ is not derivable by MV$_2$, MV$_3$ and MV$_4$. The queries were entered by several users over a three-day period as given in Figure 13. It is observed that Q$_5$ is entered daily during the three-day period and since it is derivable using MV$_2$, MV$_2$ is recognized as a materialized view with high demand. Therefore, there is a high probability that MV$_2$ will be used on day 4 as well. Contrary to this Q$_6$ was entered only on day 1 and therefore MV$_6$ was not popular among the users, thus the probability of it being used on the fourth day is low. Q$_7$ does not have a materialized view that is able to derive its result but users have entered it into the system on all three days. Therefore, we can say that Q$_7$ has been a popular query within the three-day period and there is a high probability that it will be entered on the fourth day. Based on this analysis and the predicted outcomes for day 4 the analyzer can execute the following changes to M:

- Create an MV using query Q$_7$ – on the fourth day the cost for query Q$_7$ is reduced
- Remove MV$_3$ – Best case is Q$_6$ is not entered. Worst case is Q$_6$ is required to be run directly on the DW

| On day 1 | On day 2 | On day 3 |
|---|---|---|
| User$_1$ -> enter Q$_5$ | User$_1$ -> enter Q$_5$ | User$_1$ -> enter Q$_7$ |
| User$_2$ -> enter Q$_6$ | User$_2$ -> enter Q$_7$ | User$_2$ -> enter Q$_7$ |
| User$_3$ -> enter Q$_5$ | User$_3$ -> enter Q$_7$ | User$_3$ -> enter Q$_5$ |
| User$_4$ -> enter Q$_7$ | User$_4$ -> enter Q$_5$ | User$_4$ -> enter Q$_5$ |

**Figure 13 User queries for three days**

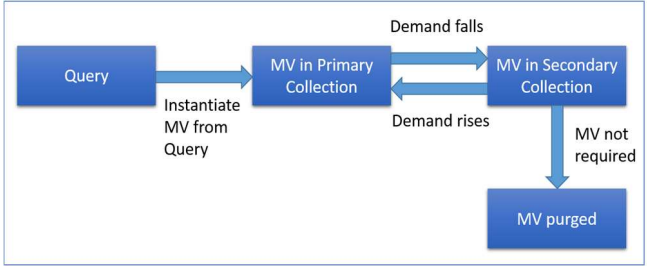

**Figure 14 Q$_{MV}$ life cycle**

This type of analysis can also be retrieved using sophisticated data mining methodologies such as clustering. However, for our approach the analysis is performed in short timespans, therefore data mining techniques would not be as effective.

In addition to creation and deletion of MVs, we also considered cases where removed Q$_{MV}$s might be in demand again. Hence, we incorporate a secondary collection of Q$_{MV}$s, denoted by N, to contain Q$_{MV}$s that are temporarily unavailable to the user due to their inactivity. Any unused Q$_{MV}$s that are removed from M and moved to N. Any Q$_{MV}$s that have an uprising demand will be moved back to M. The advantage of maintaining N aids in reducing the search space of M. Thereby it avoids iteration over unused Q$_{MV}$s when looking for a match for Q. By having unused Q$_{MV}$s in N re-creation of MVs is avoided. This is cost effective as the cost for moving a Q$_{MV}$ between the collections is less than re-creating it. A re-creation would require re-executing costly database operations such as joins, group by etc. In addition to storing inactive Q$_{MV}$s, N is configured to store new queries which are not derivable by existing Q$_{MV}$s. The query log records their usage for the analyzer. Considering the above steps, the state transitions of a Q$_{MV}$ are illustrated in Figure 14.

| Session_Id | Q$_{MV}$_or_Query_Id | Hits |
|---|---|---|
| 1 | Q$_{MV1}$ | 5 |
| 1 | Q$_{MV2}$ | 1 |
| 1 | Q$_1$ | 0 |
| 2 | Q$_{MV1}$ | 1 |
| 2 | Q$_{MV2}$ | 1 |
| 2 | Q$_1$ | 1 |

**Figure 15 Query Log Example (database table)**

*6.1.1 Record query patterns in sessions*

To perform SCV operations on the MV collection, the usage of existing MVs from both primary and secondary collections and of new queries are required to be monitored for a particular period of time. Therefore, we define this time period as a *Session*. For example, a session can last one hour, two hours or even a day etc. During this period the analyzer collects frequency of usage of each MV and new query within the session. Furthermore, in order to make accurate predictions it would be most useful to analyze several sessions and median outcome can be used by SCV operations. We identify that the length of a session and the number of sessions to take into account should be able to be varied as it depends on the DW environment. For example, if the reporting requirement is for an hourly basis, the session length can be two hours and the analyzer looks at usage data from five such sessions.

*6.1.2 Query log for recording usage of MVs and Queries*

The query log stores the number of hits for each Q$_{MV}$ and query against a session. For a particular session, the system assigns hit counters for each MV and all new queries. At the completion of a session this data is copied to a permanent query log such as a database table and the hit counters are reset to zero. The process repeats until the query log collects usage data for a given number of sessions, then the analyzer analyzes this and perform SCV operations. Once the analysis is performed, the hit counters in the log are set to zero. A sample of the query log as a database table is given below:

*6.1.3 A scenario for analyser*

Let us consider a scenario where a DDW contains the materialized views MV$_5$ and MV$_6$ in the primary collection M, and MV$_7$ and MV$_9$ in the secondary collection N. Q$_{10}$ is derivable by MV$_5$, Q$_{11}$ is derivable by MV$_6$ and Q$_{12}$ is derivable by MV$_7$. Let M = $\{(MV_5, 0), (MV_6, 0)\}$ and N = $\{(MV_7, 0), (Q_9, 0)\}$, where the second item of a pair is the hit counter for the first item. The analyzer executes after the completion of three sessions. Assume the queries are entered as depicted in Figure 16, and the snapshot of the corresponding query log is given in Figure 17. The following SCV operations take place during the analysis process:

- Remove MV$_6$ from primary collection and add to secondary collection – MV$_6$ had no hits in all sessions. It has lost its popularity.
- Remove MV$_7$ from secondary collection and add to primary collection – MV$_7$ had hits in all sessions. It has become popular again.

| Session | Queries Presented | Primary Collection | Secondary Collection |
|---|---|---|---|
| 1 | $Q_{10}, Q_9, Q_{10}, Q_{12}$ | $\{(MV_5, 2), (MV_6, 0)\}$ | $\{(MV_7, 1), (Q_9, 1)\}$ |
| Insert usage data to database table<br>Reset session -> primary = $\{(MV_5, 0), (MV_6, 0)\}$ and secondary = $\{(MV_7, 0), (Q_9, 0)\}$<br>Next session Id = 2 | | | |
| 2 | $Q_{10}, Q_{12}, Q_9$ | $\{(MV_5, 1), (MV_6, 0)\}$ | $\{(MV_7, 1), (Q_9, 1)\}$ |
| Insert usage data to database table<br>Reset session -> primary = $\{(MV_5, 0), (MV_6, 0)\}$ and secondary = $\{(MV_7, 0), (Q_9, 0)\}$<br>Next session Id = 3 | | | |
| 3 | $Q_{10}, Q_{12}, Q_9$ | $\{(MV_5, 1), (MV_6, 0)\}$ | $\{(MV_7, 1), (Q_9, 1)\}$ |
| Insert usage data to database table<br>Reset session -> primary = $\{(MV_5, 0), (MV_7, 0), (MV_8, 0)\}$ and secondary = $\{(MV_6, 0)\}$<br>($MV_8$ is created using $Q_9$)<br>Next session Id = 1 | | | |

**Figure 16 Collection of usage data over three sessions**

- Create a new $MV_8$ for $Q_9$ and store it in primary collection – $Q_9$ had hits in all sessions and we predict that a user will enter it in the next session
- Remove $Q_9$ from secondary collection

Moreover, if no hits are recorded for $MV_6$ in the next 3 sessions, it is permanently removed from the N and thereby completely discarded from the DDW as well. After performing the above actions, the analyzer begins a new session by resetting the log counters and continues to collect usage information. The above steps of the analyzer are pseudocoded in Algorithm 4.

| **Algorithm 4** Analyzer algorithm |
|---|
| **Input**: primary: primary MV collection M,<br>secondary: secondary MV collection N,<br>query_log: Query log (database table),<br>session_id: id of session,<br>session_threshold: the number of sessions to analyze |
| **Output**: None |
| 1: Reset all hit counters |
| 2: if session_id equals session_threshold then |
| 3:   for all $mv \in primary$ do |
| 4:     temp := query result set |
| 5:     sum_of_hits := integer |
| 6:     temp <- Get from log where id = mv.id |
| 7:     for all $row \in temp$ do |
| 8:       sum_of_hits = sum_of_hits + row.hits |
| 9:     end for |
| 10:    if sum_of_hits equals 0 |
| 11:      add mv to secondary |
| 12:      remove mv from primary |
| 13:    end if |
| 14:  end for |
| 15:  for all $mv \in secondary$ do |
| 16:    temp := query result set |
| 17:    count := integer |
| 18:    temp <- Get from log where id = mv.id |
| 19:    for all $row \in temp$ do |
| 20:      if row.hits >= 1 then |
| 21:        count = count + 1 |
| 22:      end if |
| 23:    end for |
| 24:    if count equals session_threshold |
| 25:      add mv to primary |
| 26:      remove mv from secondary |
| 27:    end if |
| 28:    if count equals 0 |
| 29:      remove mv from secondary |
| 30:    end if |
| 31:  end for |
| 32: truncate query_log |
| 33: session_id = 1 |
| 34: end if |

| Session_Id | MV_or_Query_Id | Hits |
|---|---|---|
| 1 | MV5 | 2 |
| 1 | MV6 | 0 |
| 1 | MV7 | 1 |
| 1 | Q9 | 1 |
| 2 | MV5 | 1 |
| 2 | MV6 | 0 |
| 2 | MV7 | 1 |
| 2 | Q9 | 1 |
| 3 | MV5 | 1 |
| 3 | MV6 | 0 |
| 3 | MV7 | 1 |
| 3 | Q9 | 1 |

**Figure 17 Query log for 3 sessions of analyzer scenario**

*6.2 DDW Change Tracker*

To provide up-to-date results using the MV collection M, we propose a tracking component to our maintenance system to identify changes in the schema or the data. DWs primarily contain historical data [16] and newly arriving data will be aggregated and added into the DW. DDWs are designed to integrate these more seamlessly than DWs. Due to this we can expect a growth in the number of records in the fact tables. Consequently, in dimensions, new records may be added or removed. The conceptual and physical schemas can change unpredictably in DDW as discussed in earlier chapters. To cater for these requirements, the *DDW change tracker* of our technique is configured to monitor the aforementioned changes in the DDW. In instances where the schema changes, new data arrives or data being removed it will produce an alert so that the MVs can be re-computed. As the $Q_{MVS}$ in M can be looked up in the hash table using the table name, the MVs affected can easily be updated by looking up the affected tables.

*6.3 Summary*

In summary, it can be stated that the maintenance of M is important as it increases the possibility of finding a match for Q. The analyzer achieves this by looking at user query patterns recorded in the query log. It is modeled to run independently from the remainder of the system and so users are able to query the DDW without interruption. DDW change tracker aims to update the content of out dated MVs by listening to changes in the DDW tables.

## 7 Experimental Setup and Evaluation

In this section, we experimentally evaluate *query matching*, *domain matching* and *MV maintenance*. We demonstrate that these three approaches are effective solutions, by making comparisons to their traditional/naïve approach counterparts.

*7.1 Experimental Setup*

*7.1.1 Machine and Database server*

The machine used for running experiments has an Intel Core i5-2450M processor with speed 2.5GHz. It contains 2 cores and 4 logical processors and supports multi-threading. This PC has a DDR3 RAM memory of 8GB with a speed of 1333GHz. Furthermore, it is installed with an Oracle 12c Release 1 Enterprise edition database server running locally, which is used to setup the data warehouse for all experiments.

*7.1.2 Simulator program*

The simulator program is written using Java. The implementation of the proposed *OLAP query tree* and the query

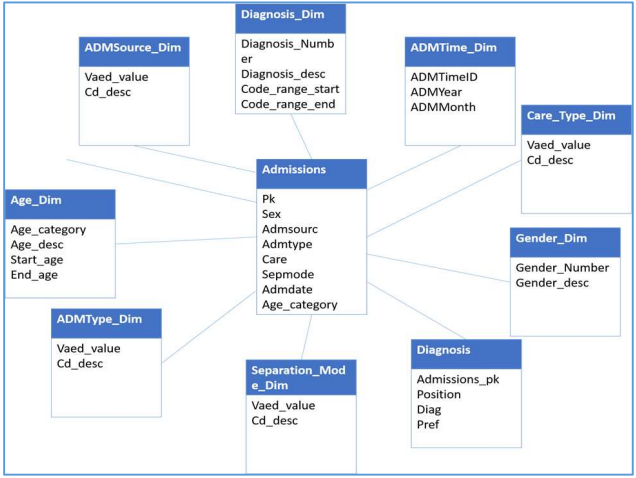

**Figure 18 Hospital admissions star schema**

| Table | Admissions | Gender_Dim | AdmSource_Dim | AdmType_Dim |
|---|---|---|---|---|
| No. of rows | 1548682 | 3 | 8 | 8 |

| Table | Age_Dim | Diagnosis | Procedures | Separation_Mode_Dim |
|---|---|---|---|---|
| No. of rows | 6 | 4893718 | 2906219 | 9 |

| Table | Diagnosis_Dim | Care_Type_Dim | AdmTime_Dim |
|---|---|---|---|
| No. of rows | 21 | 18 | 120 |

**Figure 19 Hospital admissions data warehouse statistics**

matching algorithm are contained within separate java packages. The implementation includes various setups that are configured to run the experiments. The code for this implementation can be downloaded from the following link:

https://bitbucket.org/charles_goonetilleke/frequent-query-matching-in-ddw

### 7.1.3 Dataset and Queries

**Dataset**: In order to create a data warehouse for experiments, we use a real dataset with hospital admission information, obtained from a hospital in Victoria, Australia. The data has been fully anonymized and any reference to any particular identity of individuals or events has been removed. This dataset contains information regarding the admission of patients, their diagnosis and procedures followed. In addition to that, information about the patients' gender, admission source, admission type, care type, separation mode, date of admission and age are also provided. All data have been cleaned to ensure data integrity. The star schema of the data warehouse constructed for our experiments is illustrated by Figure 19 and its statistics are given in Figure 20.

With each type of experiment, the scalability of the technique is tested, therefore the admissions table is horizontally partitioned to obtain subsets of the fact data. Subsets with incremental sizes are created in the range 100-2,000 rows with an increment of 100, in the range 1,000-20,000 rows with an increment of 1000, in the range 10,000-200,000 rows with an increment of 10,000 and in the range 100,000-1,500,000 with an increment of 100,000. These subsets are used to perform horizontal partitioning on the diagnosis table.

**Queries:** The OLAP queries chosen for experiments are given in the Appendix along with the range for the number of rows produced. The queries $Q_{13}$-$Q_{18}$ are designed to target specific costs of a query. $Q_{13}$ contains a large number of tables and joins and therefore the cost of processing the query is high with respect to the join operations. $Q_{14}$ contains a large number of *group by* attributes, hence it affects the cost of grouping a given data set. $Q_{15}$ contains several selections and these increases the cost of the selection operation of a query. $Q_{16}$ contains a large number of *joins*, *group by* attributes and *selection* conditions and therefore it produces a huge cost at query time. $Q_{17}$ and $Q_{18}$ contain special functions in *selection* conditions and *grouping by* conditions, respectively.

### 7.2 Query Matching Experiments

In order to evaluate the efficiency of *Query Matching* (QM), it must be compared against *Directly querying* (DQ) the DW. Therefore, four experiments are conducted using the following four distinct queries:

$Q_{13}$ – Contains 10 tables, 9 joins and 2 group by attributes

$Q_{14}$ – Contains 2 tables, 1 join and 8 group by attributes

$Q_{15}$ – Contains 1 table, 6 selection conditions and 1 group by attribute

$Q_{16}$ – Contains 10 tables, 9 joins, 5 selection conditions and 8 group by attributes

The simulator program is setup to contain a collection of $OQT_{MV}s$ and $OQT_{Q}s$ and the algorithm for the matching. Queries $Q_{13}$-$Q_{16}$ are used to create four $OQT_{MV}s$ in the simulator program and the corresponding materialized views are created in the DW. Four $OQT_{Q}s$ that match with the $OQT_{MV}s$ are created and it is ensured that a given $OQT_Q$ only matches with one $OQT_{MV}$ only.

Each $OQT_Q$ is submitted to the simulator and the time taken by the QM to find a match and retrieve all the rows of the MV and the time taken by the simulator to obtain all the rows from the DW for the OLAP query of the $OQT_Q$ is recorded. The simulator is configured to repeat these recordings for both cases for 15 times. We observed that the simulator internally performed some caching mechanism and therefore the time recordings for the first 1-5 runs are different than the remaining recordings. To obtain a fair result the first 5 recordings are discarded and the average of the remaining are calculated.

The time taken by the QM and DQ can be represented using the following formulas:

$T_{FM} = Time\ taken\ to\ find\ match\ using\ QM$

$T_{R_{QM}} = Time\ taken\ to\ retrieve\ rows\ for\ QM$

$T_{R_{DQ}} = Time\ taken\ to\ retrieve\ rows\ for\ DQ$

$T_O = Ti\quad taken\ for\ query\ operations\ in\ DQ$

$T_{QM} = Total\ time\ taken\ for\ QM$

$T_{DQ} = Total\ time\ taken\ for\ DQ$

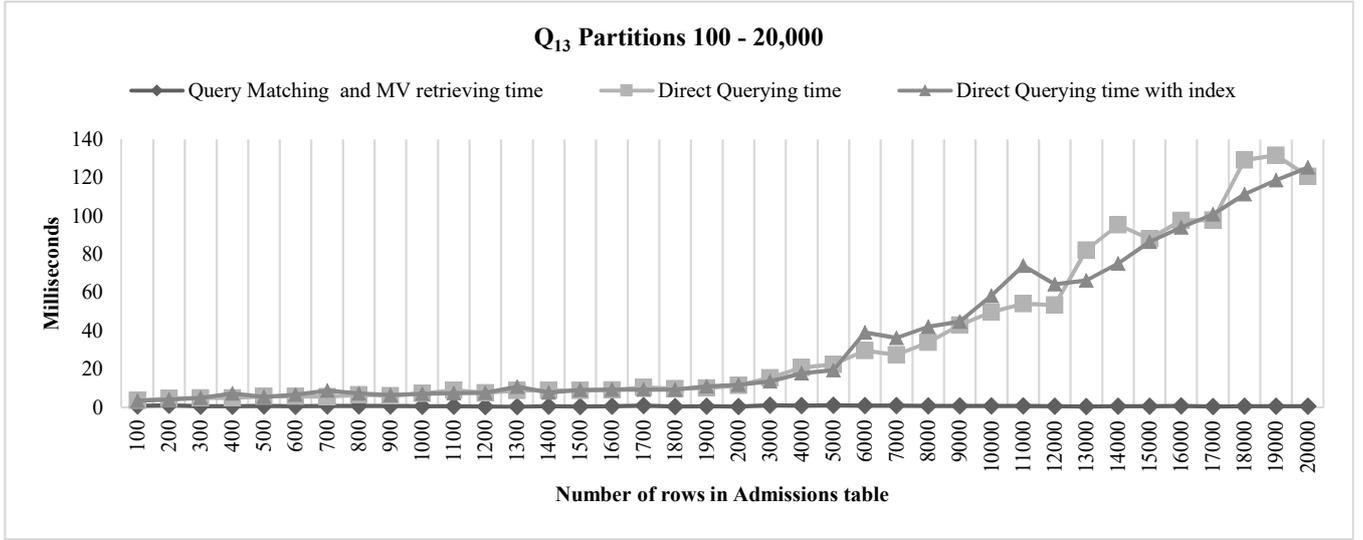

**Figure 21 Q$_{13}$ partitions 100 – 20,000**

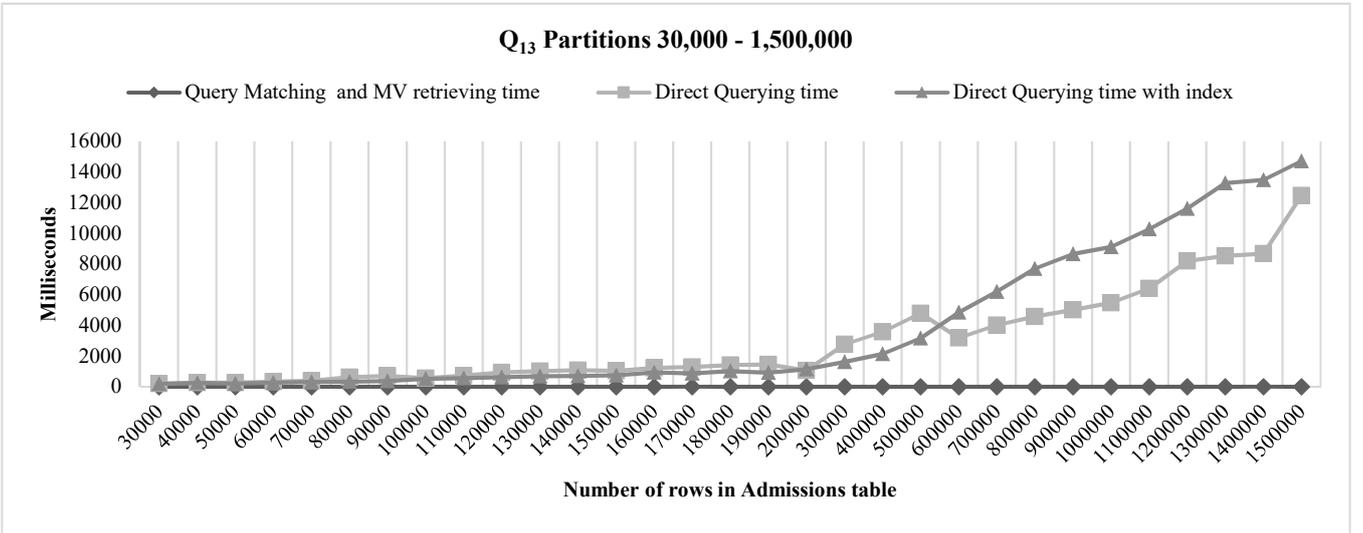

**Figure 20 Q$_{13}$ partitions 30,000 – 1,500,000**

$$T_{QM} = T_{FM} + T_{R_{QM}}$$
$$T_{DQ} = T_O + T_{R_{DQ}}$$

To collect information on the performance of QM for different sizes of the data set, each OQT$_Q$ experiment is repeated for all incremental subsets of Admissions and Diagnosis tables stated in section 7.1.3.

### 7.2.1 Results

*Q$_{13}$ – Large number of joins*

Q$_{13}$ returns a result of 6-15 rows (refer Appendix), and the MV created using it would have the same number of records. Therefore, $T_{QM}$ is small as it only performs a search to find the OQT$_{MV}$ and then retrieve 6-15 records from the DW ($T_{R_{QM}}$ would be small). As the data set gets larger, significant change in this time is not seen as the maximum number of records retrieved is limited to 15 rows. The cost of joining 10 tables is considered only at MV creation time. DQ always takes longer time than the QM as it has to perform 9 joins between 10 tables each time the query is submitted. As the data size increases, the cost of each join increases as more rows are needed to be joined, therefore $T_{DQ}$ increases exponentially.

*Q$_{14}$ – Large number of group by attributes*

For Q$_{14}$, figure 25 shows that QM performs worse than DQ. We observed that the $T_{DQ}$ is relatively less than $T_{QMT}$. Both methods retrieve the same number of rows, therefore $T_{R_{DQ}}$ and $T_{R_{QMT}}$ would be closely similar. When the algorithm does a comparison between a OQT$_Q$ and the OQT$_{MV}$, as there are several group by attributes to be compared, some extra time is taken and additionally the remaining parts of the QM adds more time as well, hence $T_{FM}$ is impacted by this; relative to this, DQ only applies the cost of the group by operation and this is less than the time taken by the QM to do query matching and so $T_O$ is bound to be less than $T_{FM}$. In Figure 24 it is observed that QM performs better than DQ. We observed that the $T_{QM}$ for datasets with 110,000 – 1,500,000 rows, does not increase considerably but $T_{DQ}$ increases by a significant amount.

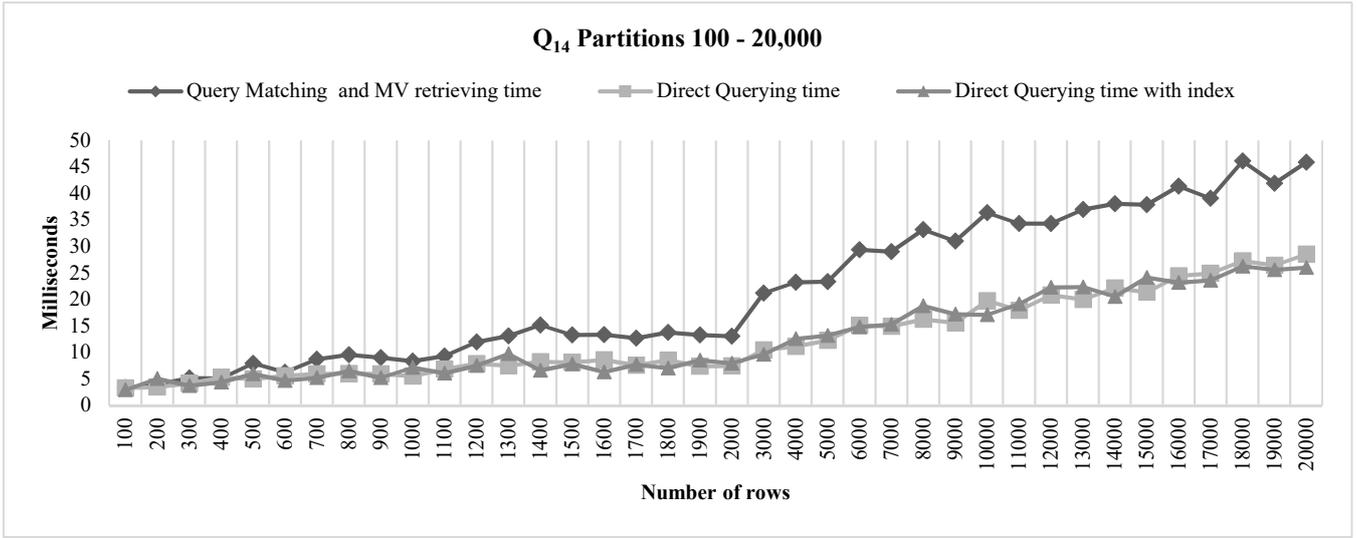

**Figure 23 Q$_{14}$ partitions 100 – 20,000**

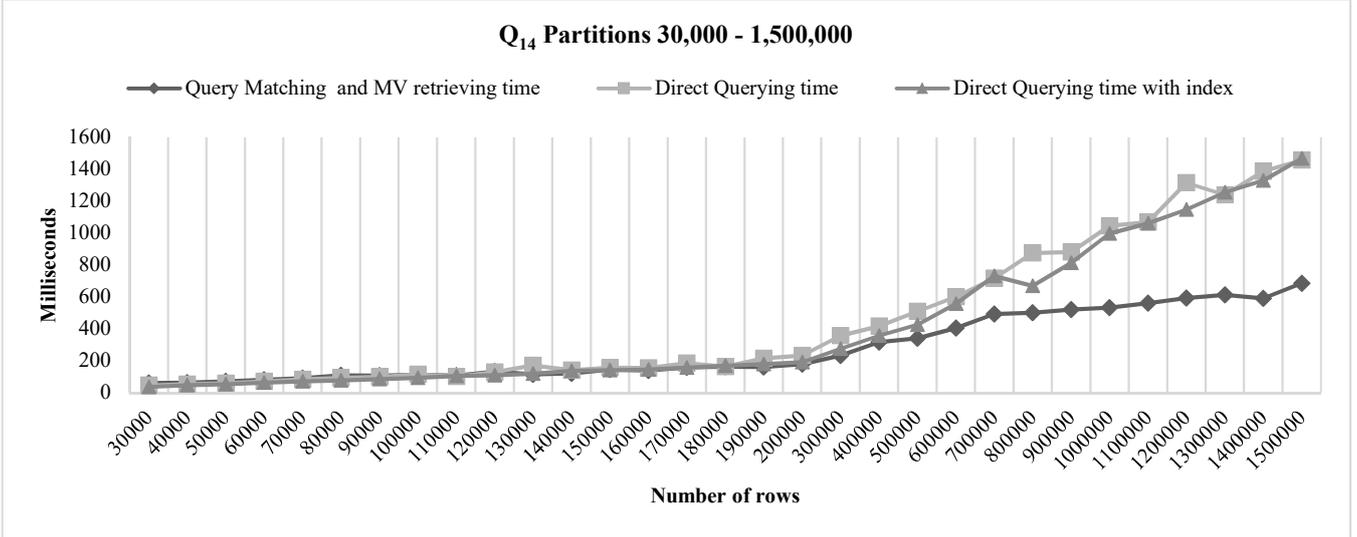

**Figure 22 Q$_{14}$ partitions 30,000 – 1,500,000**

Here $T_{RDQ}$ and $T_{RQMT}$ would remain to be closely similar but as the size of the data set grows, the group by operation produces a significant cost and this impacts on $T_O$.

*Q$_{15}$ - Large number of selection conditions*

As observed Figure 26 and Figure 27 in almost every case, $T_{DQ}$ is higher than $T_{QMT}$. The result for Q$_{15}$ would only contain 3-6 rows and therefore the total time taken by QM $T_{QMT}$ is low as $T_{FM}$ and $T_{RQMT}$ would be quite small. $T_{QMT}$ remains in the range of 0-3 milliseconds whereas $T_{DQ}$ increases by a substantial amount as the data set grows. This is due to the selection operation cost increasing with the increasing size of the data set.

*Q$_{16}$ – Large number of joins, group by and selection conditions*

In Figure 28 and Figure 29 it is observed that $T_{DQ}$ is higher than $T_{QMT}$. For Q$_{16}$ the number of rows in the result remains in the range 1-1445. Therefore, $T_{DQ}$ is higher than $T_{QMT}$, as $T_O$ would have had a substantial impact on it due to the cost of the large number of operations required to be performed. Contrary to this QM only has a small time for $T_{FM}$, therefore $T_{QMT}$ remain less than 20 milliseconds.

### 7.2.2 Efficiency Analysis

For queries Q$_{13}$, Q$_{15}$ and Q$_{16}$ performance of QM is consistently better than DQ in all sizes of data sets. $T_{RQMT}$ and $T_{RQMT}$ remains closely similar as the final result obtained by QM and DQ contains the same number of rows and as the simulator program retrieves data using its own internal DB access mechanism for both methods. Under DQ, the query operations required to find the appropriate OQT$_{MV}$ and then retrieve the result from the corresponding MV. Therefore, it can be deduced are required to be performed at run time whereas QM is only that for Q$_{13}$, Q$_{15}$ and Q$_{16}$ $T_O$ is larger than $T_{FM}$ for all data sets. Therefore, there is a considerable benefit for using QM over DQ. The only exception is for smaller data sets with queries of the type Q$_{14}$.

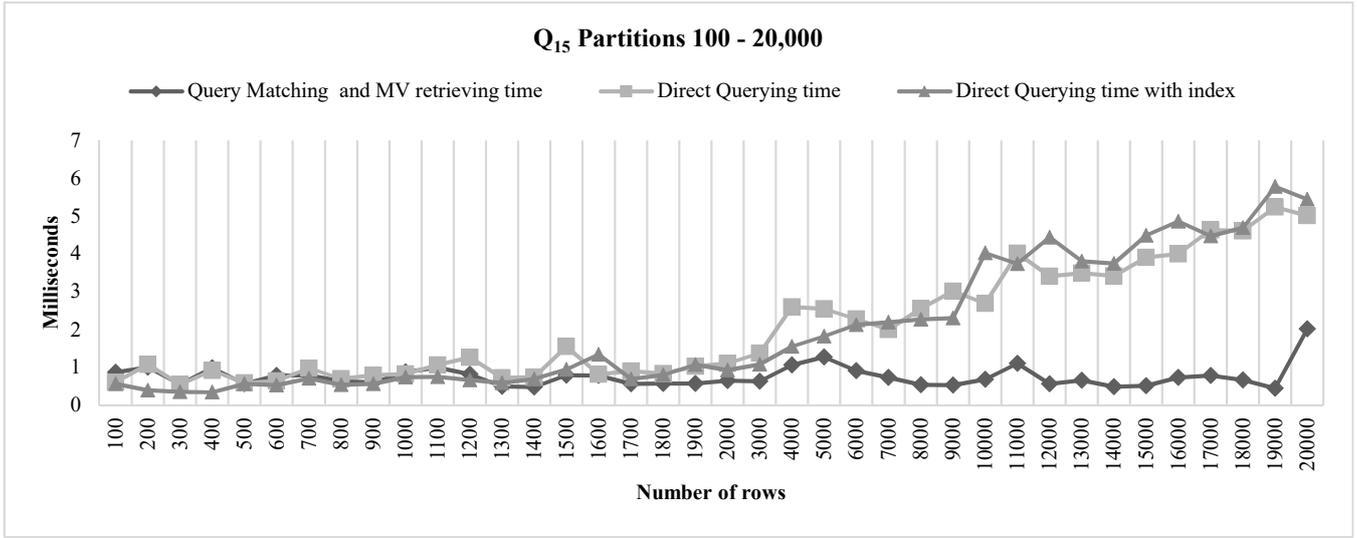

**Figure 25 $Q_{15}$ partitions 100 – 20,000**

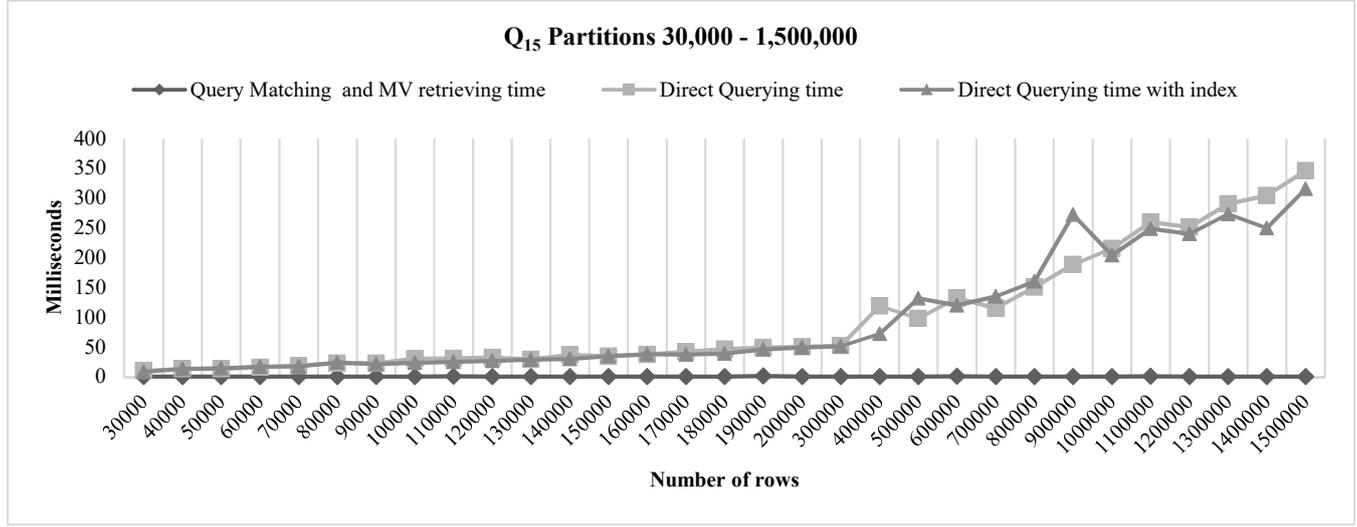

**Figure 24 $Q_{15}$ partitions 30,000 – 1,500,000**

As there are several number of group by attributes to be compared in the query matching algorithm, for data sets with less than 100,000 records, $T_{FM}$ is greater than $T_O$. But beyond 100,000 records, $T_O$ is higher than $T_{FM}$ as the grouping of >100,000 of records under 8 attributes produce a higher cost than the time taken by the algorithm to return a match.

To further verify the performance of QM, all of the above experiments are repeated with non-unique indexes on the Admissions and Diagnosis subset tables. It is observed that for $OQT_{MVS}$ made using $Q_{13}$, $Q_{15}$ and $Q_{16}$, the index did not affect the performance of DQ and so QM has a lower time for all cases. Similar trends seen between experiments performed for $Q_{14}$, without index and with index, and it is observed that DQ takes more time than QM for subsets with >100,000 records for both cases.

*7.2.3 Summary*

It can be concluded that our proposed *Query Matching* (QM) technique demonstrates better performance for most cases compared to *Directly querying* (DQ) the DW. This is also true when indexes are present. However, QM has an unsatisfactory performance than DQ for queries that contain many Group By conditions but this only occurs for tables with less than 100,000 records. In DDW environments the expected amount of data is large than this value, hence, it can be justified that this weakness of the QM does not occur in most DDWs.

*7.3 Domain Matching Experiments*

In *domain matching* (DM), we find candidates for query matching (QM) by pruning the $Q_{MV}$ collection in M. This is achieved by selecting $Q_{MVS}$ that are equivalent to Q in terms of tables. This optimization technique is further improved by using a hash table with table name as search key, for faster lookup. In order to evaluate the effectiveness of our DM, it must be compared against a trivial searching mechanism such as iterating each $Q_{MV}$ in M to find a match. The simulator program is setup to contain $OQT_{MVS}$ for queries $Q_{13}$-$Q_{17}$ and the following collections,

1. Hash table with table name as search key
2. List of $OQT_{MVS}$

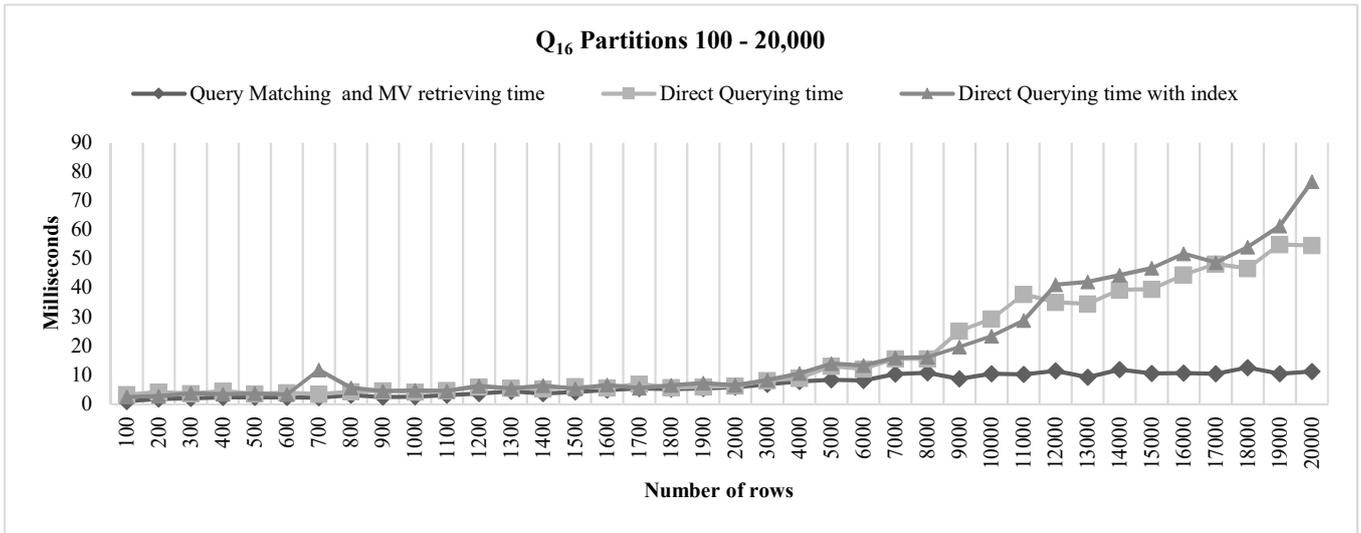

**Figure 27 $Q_{16}$ partitions 100 – 20,000**

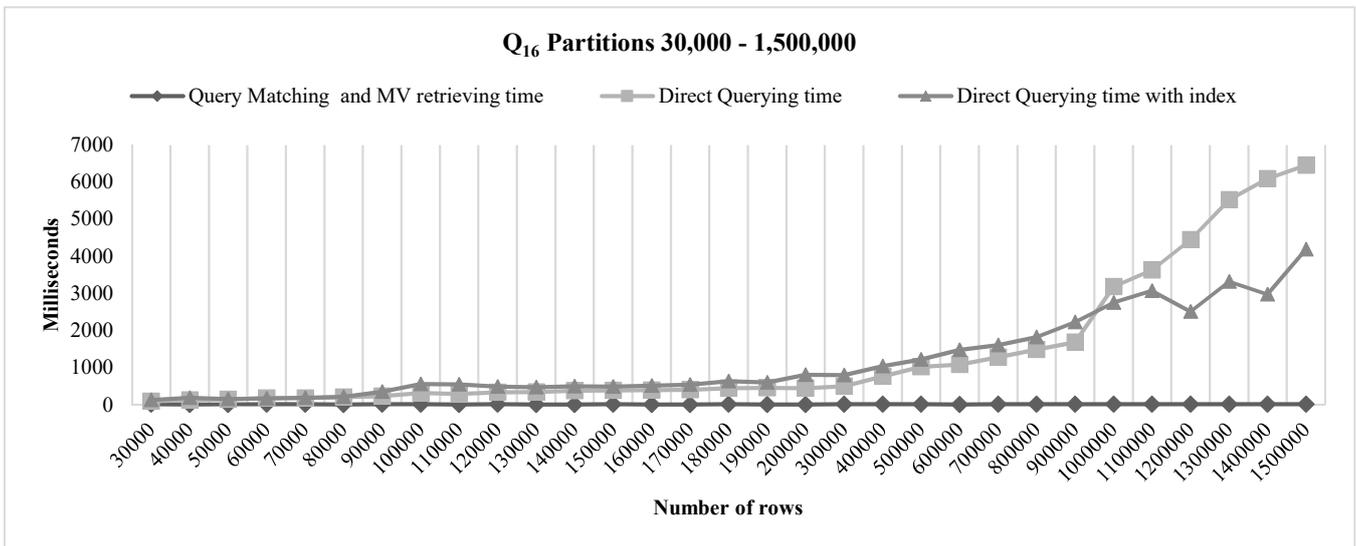

**Figure 26 $Q_{16}$ partitions 30,000 – 1,500,000**

As the Admissions is divided into 70 subsets (section 7.2.3), the program is set to loop 70 times for each subset. In each loop, it would add five $OQT_{MV}$s to each collection. Since a different subset of the Admissions table is used by each loop, five distinct $OQT_{MV}$s are added to the collections in each iteration. For example, if run 1 uses Admissions_100, it would add $OQT_{MV}$s that use Admissions_100, Diagnosis_100; Then run 2 would add $OQT_{MV}$s that use Admissions_200, Diagnosis_200 and so on. With each iteration, the number of unique tables within each collection is increased by two and the number of unique $OQT_{MV}$s is increased by 5.

Under each loop for the subsets, the program would find the time taken by the three collections to return a match for a given $OQT_Q$. To ensure a fair result is obtained for the average, the times are recorded 35 times for each collection. The first 5 records of each set are discarded and then the average time taken by each collection is calculated. This process is repeated as the number of elements in the collection are set to grow by each subset loop. $Q_{16}$ is used as the $OQT_Q$ as it contains a large number of joins, selections, group by attributes and also special functions in the selection.

### 7.3.1 Results analysis

The results of the domain matching experiments are illustrated in Figure 28 and Figure 29. The hash table with table name as search key takes the least amount of time to find the match. As the number of elements are increased from 5 to 10, our proposed domain matching technique performs significantly better in all experiments. The time taken by DM remains under 1 millisecond whereas time taken by the list increases exponentially.

### 7.3.2 Summary

In summary, it can be stated that domain matching technique is beneficial as it identifies obvious mismatches for a query Q at an earlier stage. Due to this, performance of the query matching process is improved as a subset of M is obtained. All the experiments conducted indicate the above.

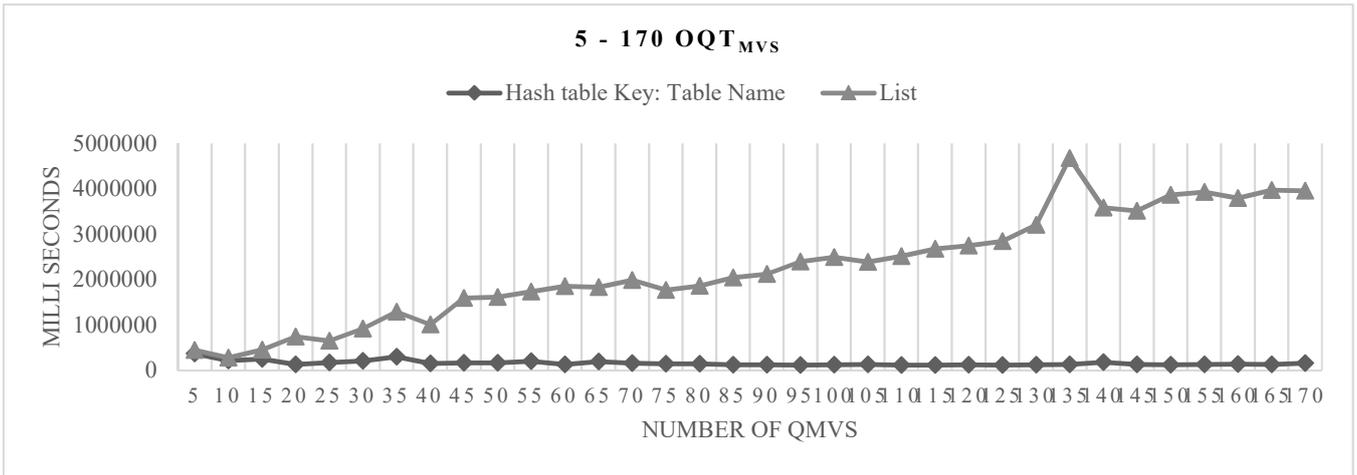

**Figure 29 Domain matching with 5-170 OQT$_{MVS}$**

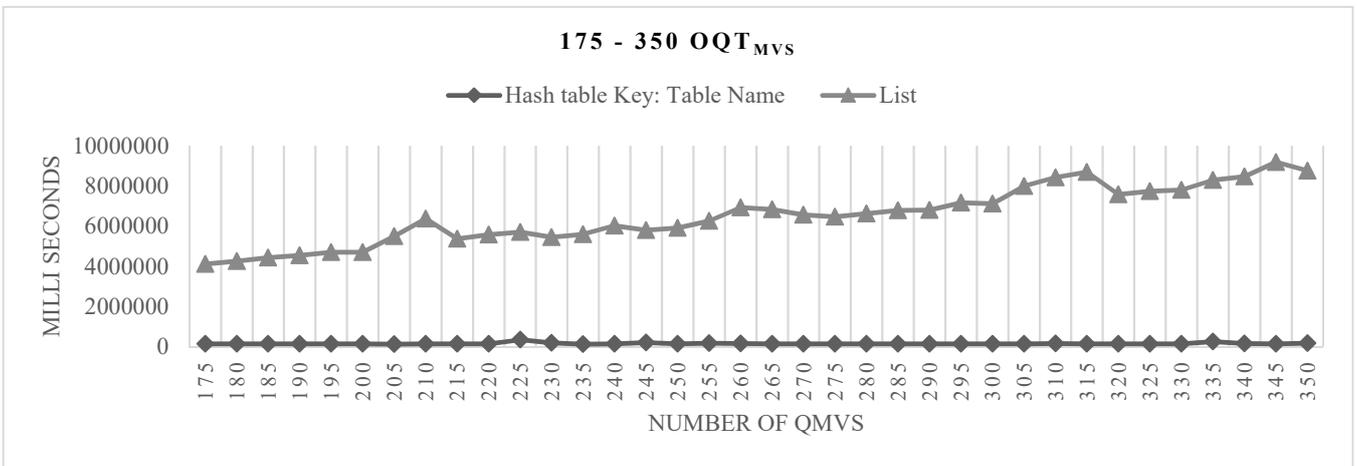

**Figure 28 Domain matching with 175-350 OQT$_{MVS}$**

*7.4 Maintenance Experiments*

The aim of MV maintenance is to maintain an intuitive set of Q$_{MVS}$ in the primary MV collection based on the frequency of queries entered by users. The *analyser component* logs the queries and the hits for each MV to perform its analysis. In our proposed analyser component, we provide flexibility to the user through the configuration of the session length and session threshold. Based on the usage of the DDW, a database administrator can change these variables to suit their needs. To make recommendations on how to choose values for session length and session threshold, we carried out a number of experiments on the analyser component. The program is setup to run experiments of 2-hour duration, in which a separate process would send queries to it at a specified interval during this period. Initially, the primary and secondary collections are empty and the program is set to contain 13 queries (variations of queries are given in Appendix). During the execution of the program, queries are to submitted by selecting one at random. We conducted 10 experiments by varying the session length, session threshold and the query frequency. The results of the experiments are summarized in Table 2.

*7.4.1   Analysis*

From experiments 1-4, the session length and query frequency is kept constant and the session threshold is incremented gradually. It is observed that the cumulative total of active Q$_{MVS}$ decreased, thereby the number of hits decreased as well. As the session threshold increases, a Q$_{MV}$'s probability of remaining active decreases. For example, cumulative total of Q$_{MVS}$ for experiment 4 is 1047, but as these Q$_{MVS}$ did not receive hits across four sessions, the analyser is not able to move them from secondary to primary collection, which resulted in the cumulative total of active Q$_{MVS}$ becoming zero. However, a high session threshold value can be used for DDWs that receive similar queries repeatedly over short periods of time. Since the average number of hits per Q$_{MV}$ per session would be higher, the pattern of hits must be analysed over a longer time span. The values for session length and session threshold also depends on the variety of queries submitted to the DDW. For environments with a broad range of queries, smaller values of session threshold are preferred. This is observed in experiments 4 and 5, where we increased the probability of similar queries being submitted by reducing the query frequency to 15 per second. As a result, the proportion of cumulative total of active Q$_{MVS}$ for session thresholds 2 and 3 increased from 8.95% to 34.02% and 2.69% to 29.83%, respectively. Furthermore, it is also observed that experiment 5 yielded a better outcome as the proportion of active Q$_{MVS}$ with hits is higher than experiment 3. For experiments 7-10, the session threshold remained constant and the session length and query frequency is varied to observe the effectiveness of the *analyser*. For experiments 9 and 10, the session length is increased and it is observed that the cumulative total of $Q_{MVS}$ decreased. Although the cumulative total of active $Q_{MVS}$ has decreased, the proportion of active $Q_{MVS}$ that received hits has increased from 62.18% to 66.76%. Thereby for longer session lengths, analyser is able to make more accurate decisions for future sessions. However, the session length must not be too high, as it could result in too many $Q_{MVS}$ being in the primary collection.

*7.4.2 Summary*

In summary, it can be stated that the *analyser component* is able to improve the usage of $Q_{MVS}$ by maintaining an intuitive collection of $Q_{MVS}$, i.e. by removing stale $Q_{MVS}$ from it and adding popular $Q_{MVS}$ to it. Throughout all experiments, it is observed that the actions of the *analyser component* did not interfere with the querying process. We experimentally determined that for a given DDW the values of session length and session threshold depends on the range of queries presented to the system and the query frequency. Based on the results we provided recommendations on how to select values for these variables.

## 8 Conclusion and Future Work

*8.1 Conclusion*

Existing *dynamic data warehouses* provide dynamicity in its design. We observed that the main problem that persists in a DDW is the mismatch between the collection of MVs and the reporting requirements of decision makers. To retrieve the result of an MV users would have to explicitly specify the MV references. This is disadvantageous as only users with prior knowledge of the MV collection would be able to utilize it. In order to improve the dynamicity of the MV collection of a DDW, we proposed a novel solution consisting of *query matching, domain matching and MV maintenance*. The query matching technique intercepts an incoming query Q and determines whether it can be answered using an existing MV in the collection. To facilitate this process, we proposed a novel structure known as *OLAP query tree*, for representing the query of the corresponding MV $OQT_{MV}$ and for representing the incoming OLAP query $OQT_Q$. The query matching algorithm receives a pair of $OQT_{MV}$ and $OQT_Q$ as inputs and determines if the result expected by $OQT_Q$ is a subset of the result in $OQT_{MV}$. As the comparison of each pair of $OQT_{MV}$ and $OQT_Q$ is an inefficient process, we proposed *domain matching* as a search space pruning technique. This optimization technique uses a hash table with table name as search key to filter $OQT_{MVS}$ that does not have same tables as Q. Query requirements of users can change over time, and to ensure a high proportion of queries are answered using MVs, we proposed a novel *MV maintenance* approach that maintains a log for every $OQT_Q$ in the collection. Furthermore, we include a *DW change tracker* to update the data in the MV collection upon change occur in the underlying data sources. Finally, we experimentally evaluate our proposed techniques against their naïve counterparts to demonstrate their effectiveness and efficiency.

*8.2 Future Work*

Our future works in this area include the enhancement of *query matching* (QM) for complex types of queries involving ROLLUP, CUBE, PARTIAL ROLL, PARTIAL CUBE functions. Further investigation of aggregate functions such as AVG, COUNT, MAX and MIN that does not show associativity as granularity decreases would be useful for improving the QM algorithm. Also, we aim to extend the *query matching* algorithm to find partial matches for an incoming query Q, such that a result can be provided by using one or more materialized views.

| Experiment | Session length (seconds) | Session threshold | Query frequency (seconds) | Cumulative total of $Q_{MVS}$ | Cumulative total of active $Q_{MVS}$ | Cumulative total of active $Q_{MVS}$ that got hits | Cumulative total of non-active $Q_{MVS}$ that got hits |
|---|---|---|---|---|---|---|---|
| EXP1 | 60 | 1 | 30 | 520 | 211 | 39 | 174 |
| EXP2 | 60 | 2 | 30 | 849 | 76 | 13 | 206 |
| EXP3 | 60 | 3 | 30 | 1229 | 33 | 5 | 259 |
| EXP4 | 60 | 4 | 30 | 1047 | 0 | 0 | 250 |
| EXP5 | 60 | 2 | 15 | 1029 | 350 | 157 | 264 |
| EXP6 | 60 | 3 | 15 | 952 | 284 | 121 | 294 |
| EXP7 | 60 | 2 | 45 | 1044 | 44 | 10 | 167 |
| EXP8 | 90 | 2 | 30 | 665 | 147 | 62 | 162 |
| EXP9 | 90 | 2 | 15 | 629 | 386 | 240 | 100 |
| EXP10 | 120 | 2 | 15 | 533 | 328 | 219 | 79 |

**Table 2 Maintenance Experiments summary**

# Appendix

The following queries were used in the Query Matching and MV Maintenance experiments:

| Q<sub>MV</sub> | Query type | Query |
|---|---|---|
| 1 | large number of joins (rows: min – 6, max - 15) | select a.sex, a.age_category, sum(1) from admissions a, diagnosis d, admsource_dim ads, admtime_dim adt, admtype_dim atp, age_dim ag, gender_dim g, care_type_dim c, separation_mode_dim s, diagnosis_dim dm where a.pk = d.admissions_pk and a.sex = g.gender_number and a.admsourc = ads.vaed_value and a.admtype = atp.vaed_value and a.care = c.vaed_value and a.sepmode = s.vaed_value and a.age_category = ag.age_category and to_char(a.admdate, 'yyyymm') = adt.admtimeid and substr(d.diag, 1, 1) = substr(dm.code_range_start, 1, 1) group by a.sex, a.age_category; |
| 2 | large number of group by conditions (rows: min – 55, max - 52637) | select a.sex, a.admsourc, a.admtype, a.care, a.sepmode, a.age_category, adt.admyear, adt.admmonth, sum(1) from admissions a, admtime_dim adt where to_char(a.admdate, 'yyyymm') = adt.admtimeid group by a.sex, a.admsourc, a.admtype, a.care, a.sepmode, a.age_category, adt.admyear, adt.admmonth; |
| 3 | large number of selection conditions (rows: min – 3, max - 6) | select a.age_category, sum(1) from admissions a where a.admsourc = 'h' and a.admtype = 'x' and a.care = '4' and a.sepmode = 'h' and to_char(a.admdate, 'yyyy') = '2007' and to_char(a.admdate, 'mm') = '12' group by a.age_category; |
| 4 | large number of joins, selection conditions and group by conditions (rows: min – 1, max - 1445) | select a.sex, a.admsourc, a.admtype, a.care, a.sepmode, a.age_category, adt.admyear, adt.admmonth, sum(1) from admissions a, diagnosis d, admsource_dim ads, admtime_dim adt, admtype_dim atp, age_dim ag, gender_dim g, care_type_dim c, separation_mode_dim s, diagnosis_dim dm where a.pk = d.admissions_pk and a.sex = g.gender_number and a.admsourc = ads.vaed_value and a.admtype = atp.vaed_value and a.care = c.vaed_value and a.sepmode = s.vaed_value and a.age_category = ag.age_category and a.admsourc = 'h' and a.admtype = 'x' and a.care = '4' and a.sepmode = 'h' and to_char(a.admdate, 'yyyymm') = adt.admtimeid and substr(d.diag, 1, 1) = substr(dm.code_range_start, 1, 1) and dm.diagnosis_desc like 'diseases%' group by a.sex, a.admsourc, a.admtype, a.care, a.sepmode, a.age_category, adt.admyear, adt.admmonth; |
| 5 | special functions in selection conditions (rows: min – 2, max - 3) | select sex, sum(1) from admissions a, diagnosis d, diagnosis_dim dm where a.pk = d.admissions_pk and substr(d.diag, 1, 1) = substr(dm.code_range_start, 1, 1) and dm.diagnosis_desc like 'diseases%' group by sex; |
| 6 | special functions in group by conditions (rows: min – 63, max - 120) | select to_char(admdate, 'yyyy'), to_char(admdate, 'mm'), sum(1) from admissions group by to_char(admdate, 'yyyy'), to_char(admdate, 'mm'); |